\documentclass[twocolumn]{aastex631}

\newcommand\hst{\textit{HST}}
\newcommand {\A}[1]{And~{\sc #1}}

\begin{document}

\title{The Hubble Space Telescope Survey of M31 Satellite Galaxies II. \\ The Star Formation Histories of Ultra-Faint Dwarf Galaxies}

\email{asavino@berkeley.edu}
\author[0000-0002-1445-4877]{Alessandro Savino}
\affiliation{Department of Astronomy, University of California, Berkeley, Berkeley, CA, 94720, USA}

\author[0000-0002-6442-6030]{Daniel R. Weisz}
\affiliation{Department of Astronomy, University of California, Berkeley, Berkeley, CA, 94720, USA}

\author[0000-0003-0605-8732]{Evan D. Skillman}
\affiliation{University of Minnesota, Minnesota Institute for Astrophysics, School of Physics and Astronomy, 116 Church Street, S.E., Minneapolis,
MN 55455, USA}

\author{Andrew Dolphin}
\affiliation{Raytheon Technologies, 1151 E. Hermans Road, Tucson, AZ 85756, USA}
\affiliation{Steward Observatory, University of Arizona, 933 N. Cherry Avenue, Tucson, AZ 85719, USA}

\author[0000-0003-0303-3855]{Andrew A. Cole}
\affiliation{School of Natural Sciences, University of Tasmania, Private Bag 37, Hobart, Tasmania 7001, Australia}

\author[0000-0002-3204-1742]{Nitya Kallivayalil}
\affiliation{Department of Astronomy, University of Virginia, 530 McCormick Road, Charlottesville, VA 22904, USA}

\author[0000-0003-0603-8942]{Andrew Wetzel}
\affiliation{Department of Physics and Astronomy, University of California, Davis, CA 95616, USA}

\author{Jay Anderson}
\affiliation{Space Telescope Science Institute, 3700 San Martin Drive, Baltimore, MD 21218, USA}

\author[0000-0003-0715-2173]{Gurtina Besla}
\affiliation{Department of Astronomy, University of Arizona, 933 North Cherry Avenue, Tucson, AZ 85721, USA}

\author[0000-0002-9604-343X]{Michael Boylan-Kolchin}
\affiliation{Department of Astronomy, The University of Texas at Austin, 2515 Speedway, Stop C1400, Austin, TX 78712, USA}

\author{Thomas M. Brown}
\affiliation{Space Telescope Science Institute, 3700 San Martin Drive, Baltimore, MD 21218, USA}

\author{James S. Bullock}
\affiliation{Department of Physics and Astronomy, University of California, Irvine, CA 92697 USA}

\author[0000-0002-1693-3265]{Michelle L.M. Collins}
\affiliation{Physics Department, University of Surrey, Guildford GU2 7XH, UK}

\author[0000-0003-1371-6019]{M. C. Cooper}
\affiliation{Department of Physics and Astronomy, University of California, Irvine, CA 92697 USA}

\author[0000-0001-6146-2645]{Alis J. Deason}
\affiliation{Institute for Computational Cosmology, Department of Physics, Durham University, Durham DH1 3LE, UK}

\author{Aaron L. Dotter}
\affiliation{Department of Physics and Astronomy, Dartmouth College, 6127 Wilder Laboratory, Hanover, NH 03755, USA}

\author{Mark Fardal}
\affiliation{Eureka Scientific, 2452 Delmer St., Suite 100, Oakland, CA 96402, USA}

\author{Annette M. N. Ferguson}
\affiliation{Institute for Astronomy, University of Edinburgh, Royal Observatory, Blackford Hill, Edinburgh, EH9 3HJ, UK}

\author[0000-0002-3122-300X]{Tobias K. Fritz}
\affiliation{ Department of Astronomy, University of Virginia, Charlottesville, 530 McCormick Road, VA 22904-4325, USA}

\author[0000-0002-7007-9725]{Marla C. Geha}
\affiliation{Department of Astronomy, Yale University, New Haven, CT 06520, USA}

\author[0000-0003-0394-8377]{Karoline M. Gilbert}

\affiliation{Space Telescope Science Institute, 3700 San Martin Drive, Baltimore, MD 21218, USA}
\affiliation{The William H. Miller III Department of Physics \& Astronomy, Bloomberg Center for Physics and Astronomy, Johns Hopkins University, 3400 N. Charles Street, Baltimore, MD 21218}

\author{Puragra Guhathakurta}
\affiliation{UCO/Lick Observatory, Department of Astronomy \& Astrophysics, University of California Santa Cruz, 1156 High Street, Santa Cruz, California 95064, USA}

\author{Rodrigo Ibata}
\affiliation{Observatoire de Strasbourg, 11, rue de l’Universite, F-67000 Strasbourg, France}

\author[0000-0002-2191-9038]{Michael J. Irwin}
\affiliation{Institute of Astronomy, University of Cambridge, Cambridge CB3 0HA, UK}

\author{Myoungwon Jeon}
\affiliation{School of Space Research, Kyung Hee University, 1732 Deogyeong-daero, Yongin-si, Gyeonggi-do 17104, Republic of Korea}

\author{Evan N. Kirby}
\affiliation{Department of Physics, University of Notre Dame, Notre Dame, IN 46556, USA}

\author[0000-0003-3081-9319]{Geraint F. Lewis}
\affiliation{Sydney Institute for Astronomy, School of Physics, A28,
The University of Sydney, NSW 2006, Australia}

\author[0000-0002-6529-8093]{Dougal Mackey}
\affiliation{Research School of Astronomy and Astrophysics, Australian National
University, Canberra 2611, ACT, Australia}

\author{Steven R. Majewski}
\affiliation{Department of Astronomy, University of Virginia, 530 McCormick Road, Charlottesville, VA 22904, USA}

\author[0000-0002-1349-202X]{Nicolas Martin}
\affiliation{Observatoire de Strasbourg, 11, rue de l’Universite, F-67000 Strasbourg, France}
\affiliation{Max-Planck-Institut fur Astronomie, K\"{o}nigstuhl 17, D-69117 Heidelberg, Germany}

\author{Alan McConnachie}
\affiliation{NRC Herzberg Astronomy and Astrophysics, 5071 West Saanich Road, Victoria, BC V9E 2E7, Canada}

\author[0000-0002-9820-1219]{Ekta Patel}
\affiliation{Department of Astronomy, University of California, Berkeley, Berkeley, CA, 94720, USA}

\author[0000-0003-0427-8387]{R. Michael Rich}
\affiliation{Department of Physics and Astronomy, UCLA, 430 Portola Plaza, Box 951547, Los Angeles, CA 90095-1547, USA}

\author[0000-0002-4733-4994]{Joshua D. Simon}
\affiliation{Observatories of the Carnegie Institution for Science, 813 Santa Barbara Street, Pasadena, CA 91101, USA}

\author[0000-0001-8368-0221]{Sangmo Tony Sohn}
\affiliation{Space Telescope Science Institute, 3700 San Martin Drive, Baltimore, MD 21218, USA}

\author{Erik J. Tollerud}
\affiliation{Space Telescope Science Institute, 3700 San Martin Drive, Baltimore, MD 21218, USA}

\author[0000-0001-7827-7825]{Roeland P. van der Marel}
\affiliation{Space Telescope Science Institute, 3700 San Martin Drive, Baltimore, MD 21218, USA}
\affiliation{Center for Astrophysical Sciences,
  The William H. Miller III Department of Physics \& Astronomy,
  Johns Hopkins University, Baltimore, MD 21218, USA}



\vspace{8mm}
\begin{abstract}

We present the lifetime star formation histories (SFHs) for six ultra-faint dwarf (UFD; $M_V>-7.0$, $ 4.9<\log_{10}({M_*(z=0)}/{M_{\odot}})<5.5$) satellite galaxies of M31 based on deep color-magnitude diagrams constructed from \textit{Hubble Space Telescope} imaging.  These are the first SFHs obtained from the oldest main sequence turn-off of UFDs outside the halo of the Milky Way (MW). We find that five UFDs formed at least 50\% of their stellar mass by $z=5$ (12.6~Gyr ago), similar to known UFDs around the MW, but that 10-40\% of their stellar mass formed  at later times. We uncover one remarkable UFD, \A{XIII}, which formed only 10\% of its stellar mass by $z=5$, and 75\% in a rapid burst at $z\sim2-3$, a result that is robust to choices of underlying stellar model and is consistent with its predominantly red horizontal branch. This ``young'' UFD is the first of its kind and indicates that not all UFDs are necessarily quenched by reionization, which is consistent with predictions from several cosmological simulations of faint dwarf galaxies. SFHs of the combined MW and M31 samples suggest reionization did not homogeneously quench UFDs. We find that the least massive MW UFDs ($M_*(z=5) \lesssim 5\times10^4 M_{\odot}$) are likely quenched by reionization, whereas more massive M31 UFDs ($M_*(z=5) \gtrsim 10^5 M_{\odot}$) may only have their star formation suppressed by reionization and quench at a later time. We discuss these findings in the context of the evolution and quenching of UFDs.

\end{abstract}

\keywords{Andromeda Galaxy -- Dwarf Galaxies -- Galaxy Evolution -- Galaxy Quenching -- Hertzsprung Russell Diagram -- Reionization}


\section{Introduction} \label{sec:intro}

Ultra-faint dwarf (UFD) galaxies around the Milky Way (MW) represent our strongest observational link between cosmic reionization and low-mass galaxy formation. Long-standing theoretical models posit that the ultra-violet (UV) background in the early Universe should prevent or suppress star formation in the lowest mass dark matter halos \citep[e.g.,][]{Efstathiou92,Bullock00,Sommerville02,Benson02,Benson03,Bovill09,Munoz09,Salvadori09,Busha10,Tumlinson10,Simpson13,Wheeler15,Jeon17}. Stellar populations of these ``fossil’’ galaxies were predicted to be largely ancient ($z>6$), but  color-magnitude diagram (CMD) analysis of classical Local Group dwarfs found virtually all of them to have substantial star formation at younger ages \citep[e.g.,][]{Tolstoy98,Gallart99,Grebel04,Monelli10a,Monelli10b,deBoer12,Weisz14,deBoer14,Skillman17,Savino19}.  The discovery of UFDs in the early 2000s \citep[e.g.,][]{Willman05a,Willman05b,Belokurov06, Sakamoto06,Belokurov07,Zucker06a,Zucker06b,Irwin07,Walsh07} and the precise determination of their star formation histories (SFHs) in the 2010s now provide the strongest evidence that reionization can stunt the formation of low-mass galaxies.  Virtually all MW UFDs with SFHs measured from the oldest main sequence turnoff (oMSTO) are consistent with a rapid quenching of their star formation between $6\lesssim z\lesssim8$ \citep{Brown12,Okamoto12,Brown14,Weisz14,Simon21,Gallart21,Sacchi21,Weisz23}, in agreement with many predictions and observational constraints on the timing and duration of reionization \citep[e.g.,][]{Fan06,Ouchi10,Pentericci11,Ono12,Robertson15,Planck16}.

Despite the uniformity of these SFHs, there remain lingering concerns about selection bias, as all these UFDs share a common accretion history associated with the dark matter halo of the MW. While theoretical and observational efforts have been dedicated to quantifying environmental effects in the evolution of the MW's UFDs \citep[e.g.,][]{Wetzel15,Rodriguez-Wimberly19,Sacchi21,Santistevan23}, the degree of synchronicity in formation and quenching among the dozens of known UFDs and the potential influence of local environment are poorly understood.

The cleanest way to alleviate these concerns is to measure the SFHs of UFDs outside the MW’s halo. Ideally, the best test would use UFDs that are unambiguously isolated, mitigating any possible effects of environment.  However, the first isolated UFD was discovered serendipitously and only within the past year \citep{Sand22}.  Instead, the only known UFDs that reside outside the MW halo, and that we can image down to the oMSTO, are satellites of M31 \citep[e.g.,][]{Martin16,Collins22,Martinez-Delgado22, McQuinn23}.  Measuring precise SFHs of these faint, distant systems ($D\sim750$~kpc) requires the excellent angular resolution and sensitivity of \hst\ and was a main science driver behind the 244 orbit \hst\ Survey of M31 Satellites program awarded in Cycle 27 (HST-GO-15902, PI: Weisz).

In this paper, we present the SFHs of six UFDs that are in the satellite system of M31. These are the first SFHs of \textit{bona fide} UFDs (i.e., $M_V > -7.5$; \citealt{Simon19}) outside the MW's halo that have been measured from their oMSTOs.  We summarize our data in \S~\ref{sec:data}, describe our methodology in \S~\ref{sec:model}, and discuss results in \S~\ref{sec:discussion}.  Throughout the paper we assume a $\Lambda$CDM cosmological model based on \citet{Planck18}.

\begin{figure*}
\plotone{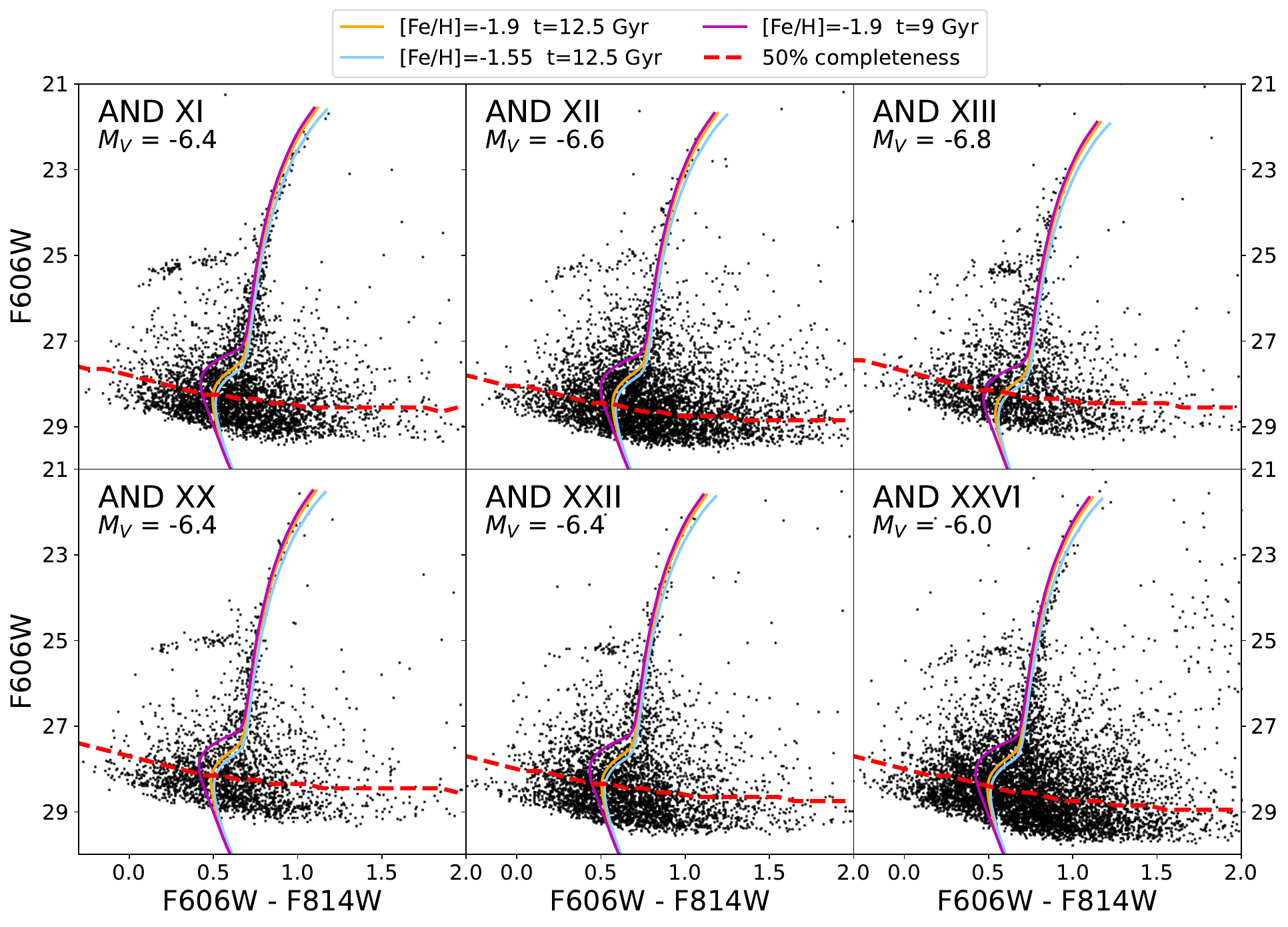}
\caption{\hst /ACS F606W vs. (F606W-F814W) CMDs for the six UFDs analyzed in this paper.   CMDs are for the central 2 half-light radii. The red dashed line marks the 50\% completeness level as determined from artifical star tests. Select BaSTI isochrones of different ages and metallicities are overplotted for reference.  Note the diversity of horizontal branch morphologies, which suggest a diversity of stellar populations, as previously found by \citet{Martin17}.  We do not include the HBs in our SFH determinations and instead use them as a sanity check on our results.}
\label{fig:CMDs}
\end{figure*}

\begin{figure*}
\plotone{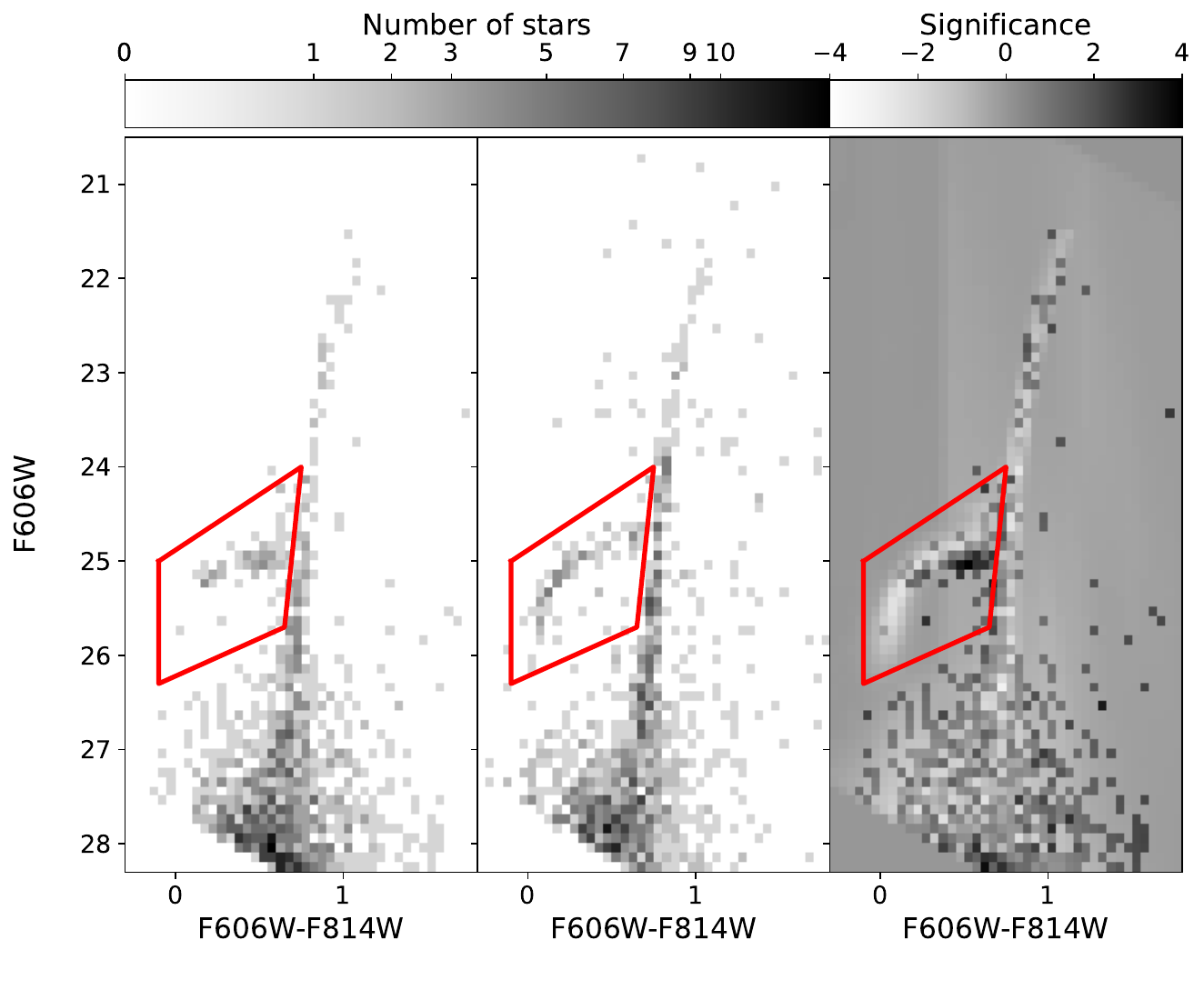}

\caption{An example of the CMD fits for \A{XX}. Shown are density maps for the observed (left panel), best-fit (center), and residual CMDs (right).  The latter is expressed in units of Poisson standard deviations. For ease of comparison, the best-fit CMD is shown as a random sampling of the underlying smooth density field. The red outline marks the HB region, which has been excluded from the fit.  The fit quality is good, with no obvious systematic structures.  The other galaxies all display similar fit qualities.}
\label{fig:Fit}
\end{figure*}

\section{Data} \label{sec:data}
We use photometric catalogs obtained as part of the \hst\ Treasury Survey of the M31 Satellite System \citep[Weisz et al., in prep.]{Savino22}. This program obtained new F606W and F814W ACS and UVIS imaging of 23 satellites of M31 without previous deep \hst\ imaging and was combined with archival \hst\ data to provide oMSTO-depth imaging for 34 known M31 satellites.

The photometry for the UFDs considered in this paper is based on a combination of new F606W and F814W \hst/ACS imaging from the \hst\ Treasury program and shallower archival imaging in the same filters taken as part of program GO-13699 \citep[PI:Martin;][]{Martin17}. We select every galaxy targeted by the survey that has an absolute magnitude $M_V>-7.0$, which leaves no ambiguity about their status as UFDs. We use absolute magnitudes from \citet{Savino22}, which are based on the PandAS structural parameters of \citet{Martin16} and are recalibrated on the basis of updated RR Lyrae distances. This selection yields six targets: \A{XI} ($M_V = -6.4\pm 0.4$, 11 orbits of imaging), \A{XII} ($M_V = -6.6\pm 0.5$, 11 orbits), \A{XIII} ($M_V = -6.8\pm 0.4$, 9 orbits), \A{XX} ($M_V = -6.4\pm 0.4$, 11 orbits), \A{XXII} ($M_V = -6.4\pm 0.4$, 18 orbits), and \A{XXVI} ($M_V = -6.0^{+0.7}_{-0.5}$, 16 orbits).

Details of the photometric reduction and catalog construction are broadly described in \citet{Savino22} and will be presented at length in the upcoming program survey paper (Weisz et al., in prep.). Here, we provide a brief summary. We reduced the ACS imaging using \texttt{DOLPHOT} \citep{Dolphin00,Dolphin16}, a well-tested point-spread function photometry package that is commonly used in the analysis of \hst\ imaging of nearby galaxies \citep[e.g.,][]{Holtzman06,Dalcanton09,McQuinn10,Monelli10a,Radburn-Smith11,Dalcanton12,Weisz14c,Williams14,Williams21}. We reduce the data using the same \texttt{DOLPHOT} set-up recommended in PHAT \citep{Williams14}, with the exception of the PSFPhotIT parameter, which we set to 2. This choice adds a second iteration to the photometric reduction, which refines the noise estimates and results in higher completeness at low S/N. 

From the resulting \texttt{DOLPHOT} catalogs, we select stars that are likely galaxy members using several criteria. First, we cull our photometric catalogs using \texttt{DOLPHOT} quality metrics. Namely we select sources with:
\begin{itemize}
    \setlength\itemsep{0.001em}
    \item $S/N \geq 4$
    \item $Sharp^2 \leq 0.2$
    \item $Crowd \leq 0.75$
    \item $Round \leq 3$
\end{itemize}
in both F606W and F814W filters. This step eliminates spurious sources, including artifacts and extended background galaxies. These selection criteria are slightly different from the PHAT recommendations \citep{Williams14}. This is because the lower stellar density in our satellite galaxies, compared to the disc of M31, allows us to apply more stringent thresholds to the values of \textit{Crowd} (the amount of contamination from neighboring sources) and \textit{Round} (the ellipticity of the source's light profile). We can therefore exclude contaminants more efficiently while retaining most stellar sources.

Second, we remove residual artifacts originating from the diffraction spikes of bright foreground stars. We do this by masking the region around all known \textit{Gaia} stars with $G<18$ that are located in or near our ACS fields. The masks are designed on a star-by-star basis to encompass the central saturated core and the visible extent of the diffraction spikes. Finally, we use the structural parameters of \citet{Martin16} to select only sources that fall within the ellipse corresponding to 2 half-light radii ($r_h$) from the photometric center of the galaxy. Our \hst\ fields are close to the photometric center in each of our targets. In most cases, however, the ACS field of view is not sufficiently large to capture the full $2r_h$ ellipse. This means that the spatial selection function varies slightly among our targets. We do, however, fully account for this variable spatial footprint whenever relevant, e.g. to calculate stellar masses (cf. \S~\ref{sec:model}).

Figure~\ref{fig:CMDs} shows the CMDs for our six UFDs. The CMDs are deep, with F606W detection limits (S/N = 4) between 29.4 and 29.8. The S/N at oMSTO ($M_{\rm F606W} \sim +4.0$) is $\sim10$. The stellar populations of these systems can be visually appreciated by the diversity of horizontal branch (HB) morphologies. We find that some of our galaxies have predominantly red HBs, confirming what was already reported from shallower \hst\ photometry \citep{DaCosta96,DaCosta00,DaCosta02,Martin17}. This is particularly the case for \A{XIII} and \A{XXII}, which almost entirely lack a blue HB.  \A{XX} and \A{XXVI} also have a high fraction of red HB stars. The red HB is typically associated with younger ages and/or higher metallicities \citep[e.g.,][]{Gratton10,Salaris13,Savino19}. We do not include the HB in our SFH fits, and instead use the HBs as a secondary check on the results which we discuss in \S~\ref{sec:extended}.

\section{CMD Modeling set up} \label{sec:model}
We model each UFD CMD using \texttt{MATCH} \citep{Dolphin02}, a commonly used software routine that recovers the system SFH by forward modeling the CMD with a combination of simple stellar population models. Details on the CMD fitting methodology are provided in the original papers and in many other nearby galaxy studies \citep[e.g.,][]{Weisz14, McQuinn15}. Here, we provide  specific details used for modeling our sample.

We use a Kroupa initial mass function \citep{Kroupa01}, normalized between 0.08~$M_{\odot}$ and 120~$M_{\odot}$, and an unresolved binary fraction of 0.35, with secondary masses drawn from a uniform mass ratio distribution. The binary fraction in UFD galaxies is poorly known. However, because our CMDs do not extend significantly below the oMSTO, this parameter as very little impact. To verify this, we have repeated our analysis assuming binary fractions of 0.5 and 0.7, and obtained virtually indistinguishable results (more details are available in the Appendix). We assume homogeneous RR Lyrae based distances, anchored to Gaia eDR3, from \citet{Savino22} and foreground extinction from \citet{Green19}.  Our adopted distance and extinction values are listed in Tab.~\ref{tab:SFH}. 

We adopt the BaSTI scaled-Solar stellar models \citep{Hidalgo18}. While many galaxies in this mass range exhibit various degrees of $\alpha$-enhancement \citep[e.g.,][]{Simon19}, this generally has a minor impact on the broadband filters, and scaled-Solar models are used in many SFH studies \citep[e.g.,][]{Monelli10a,Skillman17}. The main difference is a zero-point offset in the recovered [Fe/H] values \citep{Salaris93,Cassisi04}. We show the effects of $\alpha$-enhanced vs scaled-Solar mixtures in Appendix~\ref{App:models} and find negligible differences in the recovered SFHs. The adequacy of BaSTI scaled-Solar models is further illustrated in Appendix~\ref{App:depth}, in which we show that the CMD of te metal-poor MW globular cluster M92 is well-mathced by the scaled-Solar isochrone of the appropriate age and metallicity.

We use a grid of simple stellar populations over an age range of $7.50<\log_{10}(t/yr)<10.15$ with a 0.05 dex resolution and metallicities that span $\rm -3.0<[Fe/H]<0$, with a 0.1 dex resolution. We adopt a physically motivated prior on the age-metallicity relationship consistent with several past studies \citep[e.g.,][]{Weisz14, Skillman17}.  Specifically, we require the metallicity to increase monotonically with time, with a modest dispersion allowed at each age (0.15 dex).  This helps mitigate the age-metallicity degeneracy at the oMSTO, which is affected by the modest S/N of the data, the limited temperature sensitivity of F606W-F814W,  and the paucity of stars on the CMDs of these faint galaxies. This prior only requires metallicity to increase with time but puts no constraints on the metallicity values themselves. Additional details about the metallicity of our targets are provided in \S~\ref{sec:extended} and further tests on the effect of this assumption are detailed in the Appendix.



\begin{table*}[t]
    \centering
    \begin{tabular}{lcclccccccc}
    \toprule
        ID &$(m-M)_0$& E(B-V)&$M_V$&$M_*(z=0)$ &$M_*(z=5)$ & $\tau_{50}$& $\tau_{80}$& $\tau_{90}$& $\tau_{20-80}$&$\tau_{20-90}$\\
        &&&&$10^5 M_{\odot}$&$10^5 M_{\odot}$&Gyr&Gyr&Gyr&Gyr&Gyr\\
        \toprule
        
        \A{XI} & $24.38\pm 0.07$ & $0.09\pm 0.03$&$-6.4\pm0.4$&$1.4_{-0.3}^{+0.3}$&$1.5_{-0.8}^{+1.0}$&$13.2_{-0.6}^{+0.9}$& $12.7_{-0.1}^{+1.4}$& $9.8_{-1.6}^{+2.8}$ & $1.0_{-1.0}^{+0.4}$&$1.2_{-1.2}^{+4.4}$\\
        
        \A{XII} & $24.28\pm 0.08$& $0.17\pm 0.03$&$-6.6\pm0.5$&$2.6_{-0.9}^{+2.0}$&$2.6_{-1.2}^{+3.0}$&$13.2_{-0.6}^{+1.0}$&$12.6_{-1.7}^{+1.5}$ &$6.8_{-0.6}^{+1.4}$ & $1.1_{-1.1}^{+1.8}$&$7.2_{-2.1}^{+0.5}$\\
        
        \A{XIII}& $24.57\pm 0.07$& $0.14\pm 0.03$&$-6.8\pm0.4$&$1.1_{-0.2}^{+0.4}$&$0.1_{-0.1}^{+0.2}$&$10.6_{-0.6}^{+0.6}$&  $10.1_{-0.1}^{+1.1}$ &$5.9_{-0.4}^{+0.5}$ & $0.9_{-0.9}^{+0.2}$&$5.6_{-1.4}^{+0.2}$\\
        
        \A{XX} & $24.35\pm 0.08$& $0.08\pm 0.03$&$-6.4\pm0.4$&$1.0_{-0.2}^{+0.2}$&$0.9_{-0.5}^{+0.6}$&$13.1_{-0.5}^{+1.0}$& $12.3_{-1.7}^{+0.4}$ &$9.1_{-0.8}^{+1.6}$ & $1.1_{-1.1}^{+2.1}$&$5.1_{-2.3}^{+0.5}$\\
        
        \A{XXII} & $24.39\pm 0.07$& $0.10\pm 0.03$&$-6.4\pm0.4$&$1.1_{-0.04}^{+0.5}$&$0.9_{-0.2}^{+0.4}$&$13.0_{-0.4}^{+1.1}$& $10.6_{-0.6}^{+1.2}$ &$8.1_{-2.7}^{+2.1}$ & $3.1_{-1.6}^{+0.8}$&$3.6_{-1.1}^{+4.8}$\\
        
        \A{XXVI} & $24.48\pm 0.07$& $0.09\pm 0.03$&$-6.0^{+0.7}_{-0.5}$&$1.8_{-0.1}^{+1.3}$&$1.8_{-0.7}^{+1.7}$&$13.1_{-0.5}^{+1.0}$&$12.0_{-1.0}^{+0.7}$ &$5.7_{-0.3}^{+1.0}$ & $1.1_{-1.1}^{+1.7}$&$7.3_{-1.2}^{+1.2}$\\
        \toprule
    \end{tabular}
    \caption{Adopted distance and reddening values for our CMD modeling, literature absolute magnitudes, and stellar masses and star formation timescales inferred from our analysis of the six UFDs.}
    \label{tab:SFH}
\end{table*}

We model observational effects (photometric errors and incompleteness) for each galaxy using $\sim5\times10^5$ artificial star tests (ASTs).  The ASTs are distributed uniformly in color and magnitude over the CMD and spatially distributed following a 2D exponential density profile that is taken from \citet{Martin16}. The ASTs are injected into the images and recovered with the same reduction setup described in \S~\ref{sec:data}.

We model the CMD from $\sim2$ magnitudes above the brightest observed red giant branch stars down to the observed magnitude that corresponds to a 50\% completeness level (28.3-28.6 in F606W and 27.6-27.9 in F814W). We use a CMD bin size of $0.05\times 0.1$ mag in color and magnitude, respectively. We exclude the HB region from the fit due to uncertainties in the models of HB stars and to the availability of more reliable age and metallicity sensitive MSTO stars in our photometry \citep[e.g.,][]{Gallart05}. In our models, we include MW foreground stars using the models from \citet{deJong10}.

\texttt{MATCH} maximizes a Poisson likelihood function to find the most likely SFH that describes the observed CMD.  We then calculate random uncertainties in the solution using the methodology described in \citet{Dolphin13}, which is based on Hamiltonian Monte Carlo sampling of the solution parameter space. We also estimate the size of systematic uncertainties due to the stellar models, using the methodology of \citet{Dolphin12}. This technique introduces perturbations in the luminosity and temperature of the reference stellar models, to simulate uncertainties in the stellar evolution parameters. We also explore the effects of stellar evolution models on our fits in Appendix~\ref{App:models}, and find they are consistent with our computed systematic uncertainties.

\begin{figure*}[t]
\plotone{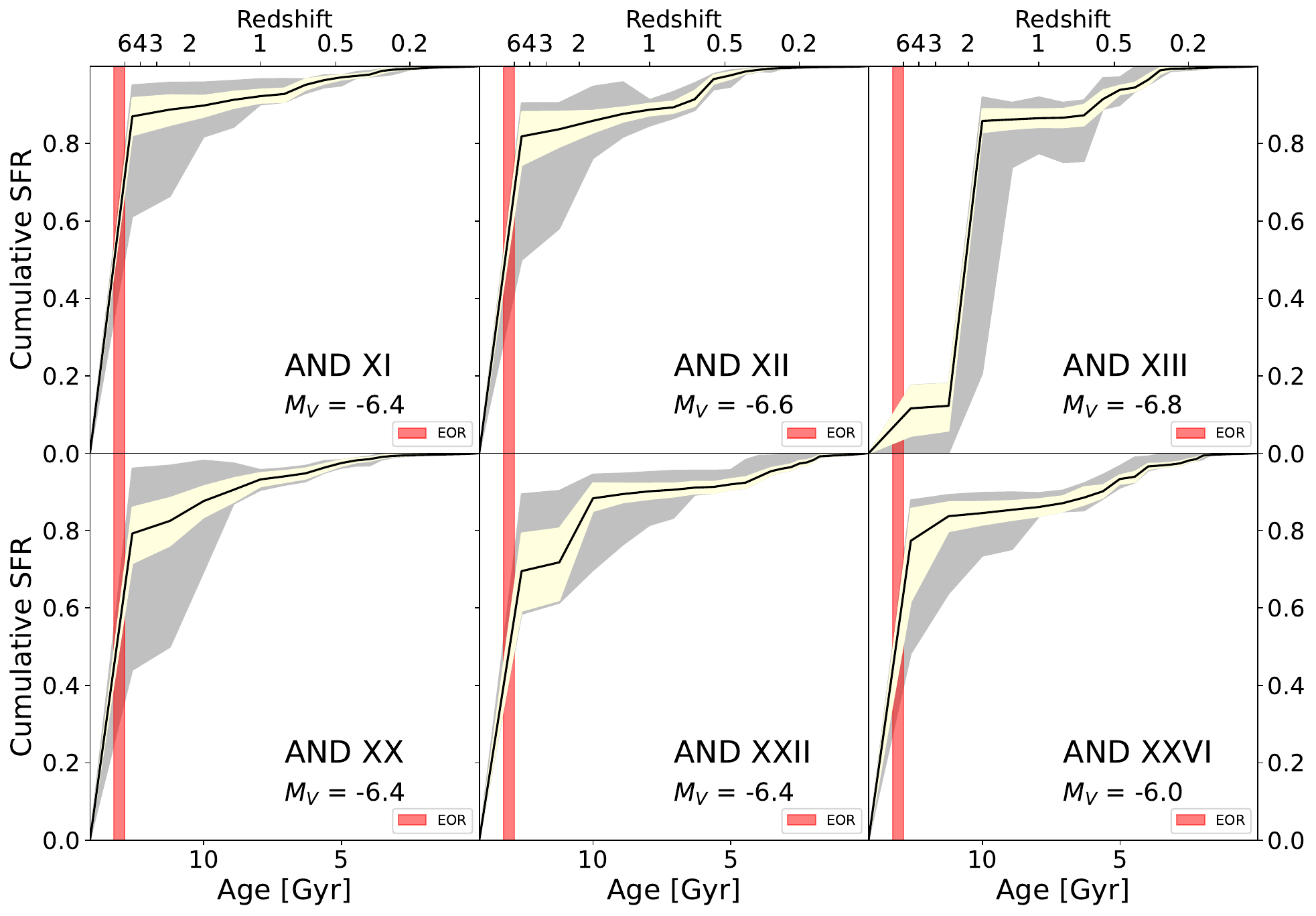}

\caption{The lifetime cumulative SFHs of the six UFDs analyzed in this paper. The black line shows the best fit SFH, the yellow shading reflects random uncertainties, and the gray shading indicates the inclusion of systematic uncertainties from stellar models. The red shaded region marks the approximate period of cosmic reionization \citep[$6\lesssim z\lesssim 9$,][]{Robertson22}. Though the majority of these galaxies have predominantly ancient stellar populations, some formed 10-40\% of their stellar mass post-reionization.  \A{XIII} is a ``young'' UFD, the first known of its kind.}
\label{fig:SFHs}
\end{figure*}


As an example for our modeling procedure, Fig.~\ref{fig:Fit} shows the observed CMD Hess diagram, the best fit CMD model and the fit residuals for one of our targets (\A{XX}). Aside from a minor discrepancy at the level of the HB (which was excluded from the fit; red box), there is good agreement between the observed and model CMD; no notable structure is visible in the residuals. The fits for the other five galaxies are of similar quality.

We use the present day and early Universe stellar masses as measured from CMD modeling as part of our analysis.  To get the total stellar mass we first integrate the SFH, which yields the birth mass of the stellar population in our \hst\ field. The present-day stellar mass in our \hst\ field is then derived by subtracting the amount of mass lost due to stellar evolution. We calculate this factor using the stellar population framework of \citet{Conroy09}, assuming a 13~Gyr stellar population with $\rm[Fe/H]=-2.0$ (cf.~\S~\ref{sec:discussion}). The resulting mass loss is 42.5\% of the stellar population birth mass. For old stellar populations, this value has very little dependence on the SFH. For stellar populations with $8<t<14$~Gyr, and $\rm-3.0<[Fe/H]<-1.0$, this factor changes by less than 2\%. Then, we must account for the limited spatial coverage of the CMDs (due to the ACS field of view and the catalog spatial cuts).  Under the assumption that the stellar mass-to-light ratio does not vary significantly across the galaxy, we use the morphological parameters from \citet{Martin16} to estimate what fraction of the stellar mass falls outside our footprint. This fraction varies from 15\% (\A{XX}) to 55\% (\A{XII}). We then obtain the total present day stellar mass for our UFDs.

The stellar mass uncertainties are calculated using a Monte Carlo approach. We obtain 5000 random realization of the stellar mass by sampling from the SFH uncertainties and the reported uncertainties in the \citet{Martin16} structural parameters. We then use the $16^{th}$ and $84^{th}$ percentile of the stellar mass distribution to estimate our confidence interval.

We also estimate the stellar mass of the galaxies at $z=5$ using a similar procedure. In this case, we use the total star formation in our oldest age bin to calculate the birth stellar mass. Because we have no constraints on when star formation occurred within our oldest time bin, stellar evolution mass loss is more uncertain. We include this effect in our Monte Carlo samples by drawing a stellar population birth time from a uniform $10.10\le \log_{10}(t)\le 10.15$ probability distribution and calculating the corresponding mass loss at $z=5$. The median mass loss we obtain with this method is 34\%.  The resulting stellar masses of the galaxies at $z=0$ and $z=5$ are listed in Table~\ref{tab:SFH}. Because star formation causes stellar mass to increase over time and stellar evolution mass loss causes a decrease in stellar mass, the inferred stellar mass at $z=5$ can be higher or lower than what measured at $z=0$, depending on the interplay between these two factors.


\section{Results and Discussion} \label{sec:discussion}

\subsection{Lifetime SFHs} \label{sec:SFHs}
Figure~\ref{fig:SFHs} shows the lifetime cumulative SFHs for our six UFDs. We focus on the cumulative SFHs because they provide a robust statistical treatment of the uncertainties, that includes the covariance among different star formation bins. However, for completeness, we also provide the instantaneous star formation rates in Appendix~\ref{App:SFH}. Five of the UFDs in our sample (\A{XI}, \A{XII}, \A{XX}, \A{XXII} and \A{XXVI}) have a prominent episode of star formation in the earliest time bin ($t>12.6$ Gyr), which drastically decreased in intensity by $z\sim5$. Taken at face value, the best-fit SFHs indicate that all five galaxies formed $\gtrsim50$\% of their stellar mass in the oldest time bin. This initial strong episode of star formation is followed by varying degrees of star formation at later times ranging from $\sim15\%$ (\A{XI}) to $\sim 35\%$ (\A{XXII}) of the total stellar mass. The significance of this SFH tail, and its fidelity to the true evolution of our targets will be further discussed in \S~\ref{sec:extended}.

\begin{figure}

\includegraphics[width=0.5\textwidth]{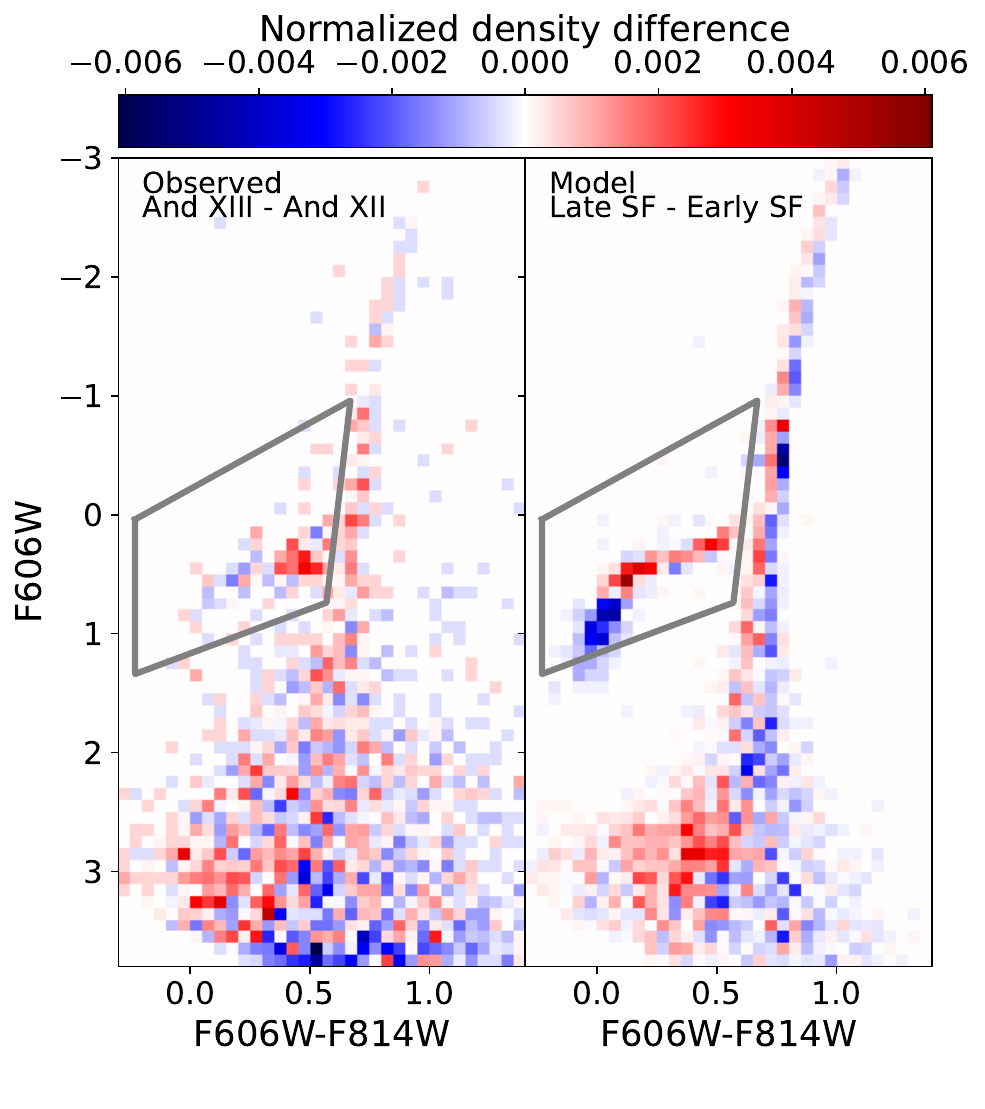}

\caption{Left: Difference between the normalized density distribution of stars in the CMD of \A{XIII} and that in the CMD of \A{XII}. Red pixels denote an excess of stars in the CMD of \A{XIII}, while blue pixels denote a deficiency. The gray box shows the HB region, which was excluded from the CMD fit. Right: Difference between the normalized density distribution of stars in a ``Late" star forming population model (average stellar age of $\sim 10.6$ Gyr) and an ``Early" star forming population model (average stellar age of $\sim 13.5$ Gyr). Compared to \A{XII}, \A{XIII} has a markedly redder HB and a higher fraction of bright MSTO stars. As supported by the models, this is a sign of a comparatively younger stellar population.}
\label{fig:Differential}
\end{figure}

\begin{figure*}

\includegraphics[width=\textwidth]{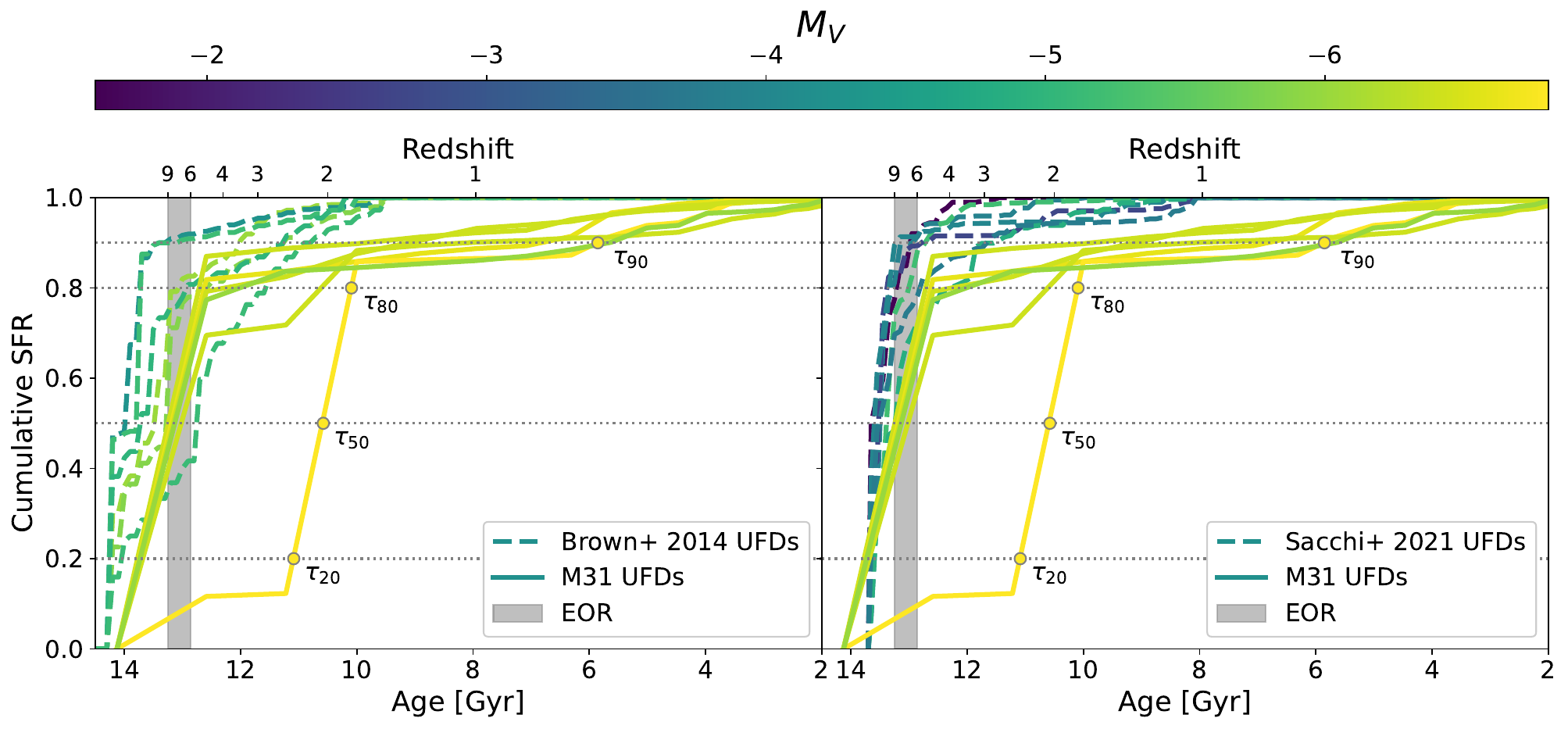}

\caption{Lifetime cumulative SFHs for our six M31 UFDs (solid lines), compared to those of MW UFD satellites (dashed lines) from the \citet{Brown14} sample (left panel) and the \citet{Sacchi21} sample (right panel), which contains galaxies typically 2 to 3 magnitudes less luminous. The SFHs are color-coded by present-day luminosity. The shaded grey region delineates the epoch of reionization. The dotted lines and circular symbols illustrate the definition of $\tau_{X}$ for one of our galaxies (\A{XIII}).}
\label{fig:SFHsample}
\end{figure*}

\A{XIII} exhibits a very different SFH.  From the best-fit SFH, we see that only 10\% of its stellar mass formed in the oldest time bin, followed by a $\sim$1~Gyr quiescent period. The galaxy then experienced a strong burst of star formation beginning 11 Gyr ago, in which it formed 75\% of its total stellar mass in the span of 1~Gyr.  

As we discuss below, the SFH of \A{XIII} is unlike any other known UFD. Its unusual nature, and the observational challenges linked to this target (being the most distant system in our sample, one of the most extincted, and with relatively shallow photometry), motivated a number of tests for robustness (e.g., sensitivity to adopted stellar model, distance/extinction variations), all of which confirm this ``young'' UFD scenario. These tests are illustrated in the Appendices~\ref{App:models}, \ref{App:Dgrid}, \ref{App:depth}, and \ref{App:Code} of this paper and reveal that, under virtually any assumption, no more than 40\% of the total stellar mass of \A{XIII} could have formed in the oldest time bin.

Figure~\ref{fig:Differential} illustrates the features in \A{XIII}'s CMD that are linked to a younger age. We have taken the Hess diagram of \A{XIII}'s CMD, removed the effect of distance and reddening, and normalized its integral to 1, effectively deriving a probability density distribution of its stars on the CMD. We did the same for the Hess diagram of \A{XII}, which has 80\% of its star formation in the oldest time bin. We show the difference between the two density maps (left panel). Because the two photometric catalogs have very similar depth and completeness, differences in the density distribution are intrinsic to the stellar population properties.

For comparison, we have also constructed a similar map (right panel) showing the difference between two stellar population models. The first one is an ``Early" star forming model, that only has star formation in the oldest time bin ($10.10<log(t)<10.14$, with an average age of 13.5 Gyr), while the second is a ``Late" star forming model, in which the star formation happens in our third time bin ($10.0<log(t)<10.05$, with an average age of 10.6 Gyr). The metallicities of the Early and Late models have been set to match the best-fit for \A{XII} ($[M/H]\sim-1.75$) and \A{XIII} ($[M/H]\sim-2.0$), respectively.

As shown in Fig.~\ref{fig:Differential}, the MSTO of \A{XIII} is significantly brighter and bluer than that of \A{XII}. This guides our fit to substantially younger ages. Although not included in the CMD fit, Fig.~\ref{fig:Differential} also highlights the much redder color of \A{XIII}'s HB compared to \A{XII}, which is compatible with the younger SFH solution. In fact, both features (brighter MSTO and redder HB) are visible in the Late-versus-Early model comparison, although the models predict bluer HBs compared to observations, due to long-standing limitations in stellar mass-loss models \citep[e.g.][]{Gratton10}.

In principle, a higher fraction of ancient stars could exist in \A{XIII} outside of our \hst\ field of view. Stellar population gradients are known to be widespread in nearby classical dwarfs \citep[e.g.,][]{Tolstoy04,Battaglia08,Vargas14,Savino19}, with young metal-rich stars being more centrally concentrated than older, more metal-poor, stars. It is currently unclear whether these gradients exist in UFDs but, if they did, they would affect the fraction of young stars in our central \hst\ field. However,  the small size of \A{XIII} \citep[$r_h\sim 0.8\arcmin$,][]{Martin16} means that our \hst\ field contains $\sim 80\%$ of \A{XIII}'s total light. Under the extreme assumption that every star outside our field of view was ancient, our SFH solution would still allow for no more than 30\% of the total stellar mass to have formed before 12.6 Gyr ago.

\subsection{Comparison with MW UFDs} \label{sec:MW}

To place our findings into a broader context, Fig.~\ref{fig:SFHsample} shows our best-fit SFHs, along side similar measurements for the 13 MW UFD satellites that have published SFHs (\citealt{Brown14}, \citealt{Sacchi21}). The SFHs are color-coded according to the galaxy luminosity. 

Figure~\ref{fig:SFHsample} shows two clear trends. First, 16 of the 19 galaxies in the combined sample formed $\sim$80\% of their stellar mass by $z\ge5$.  Two of the exceptions are Ursa~Major I from \citet{Brown14} and \A{XXII} from this study, which formed 60\% and 65\% of their stellar mass by $z=5$.  The other exception is \A{XIII}.  Second, in general, the lowest-luminosity galaxies ($M_V\lesssim -6$) quench earlier than more massive UFDs. This trend is clearly evident in the comparison with the \citet{Sacchi21} sample, which is composed of particularly faint galaxies, but is also appreciable in the more massive sample of \citet{Brown14}. Importantly, because the lowest-luminosity systems are typically MW UFDs and the higher-luminosity systems are M31 satellites, it is possible that this is not exclusively a stellar mass trend, but could be related to differences in the early environment of the two samples.  We revisit this point later in the discussion.

\begin{figure}

\includegraphics[width=0.5\textwidth]{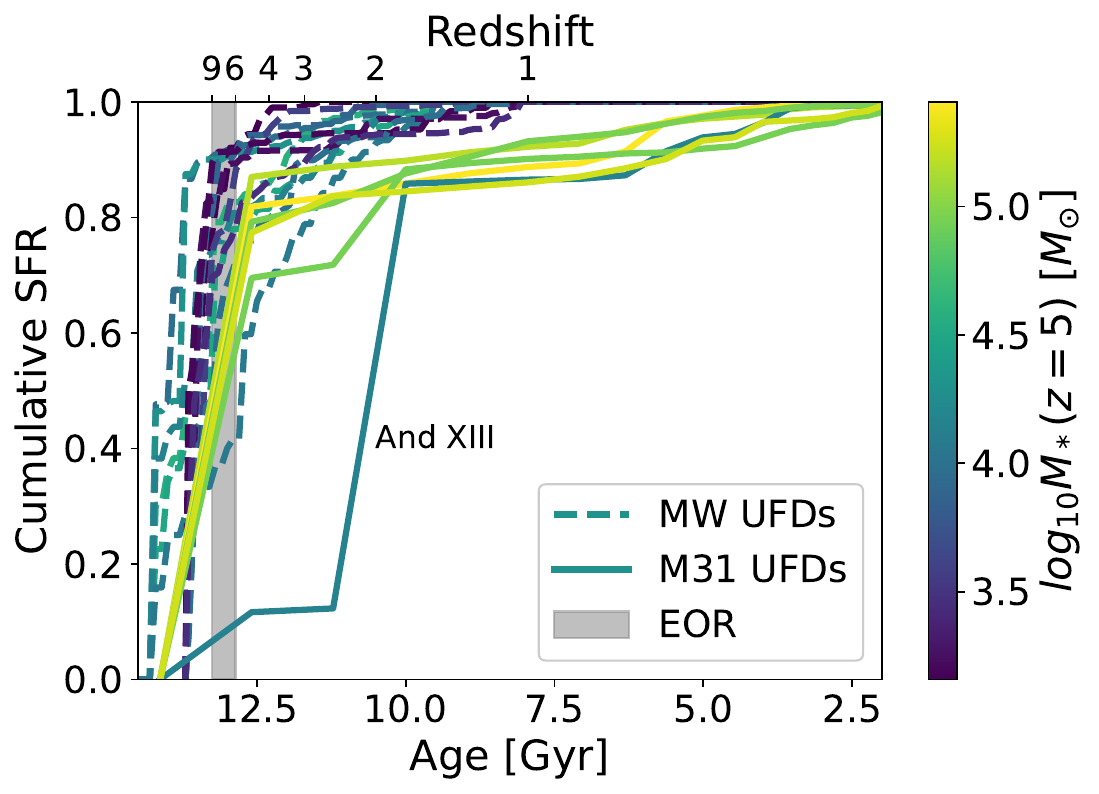}

\caption{Lifetime cumulative SFHs for our six M31 UFDs (solid lines), compared to those of MW UFD satellites (dashed lines), color-coded by total stellar mass at $z=5$. The shaded grey region delineates the epoch of reionization. The SFH of \A{XIII} is highlighted in the figure.}
\label{fig:SFHz5}
\end{figure}

Figure~\ref{fig:SFHz5} shows the same comparison of Fig.~\ref{fig:SFHsample}, color-coding the SFHs by stellar mass at $z=5$. Because of the mostly ancient SFHs, the present-day luminosity roughly traces the high-redshift stellar mass, preserving the trend observed in Fig.~\ref{fig:SFHsample}. However, \A{XIII} stands out in this comparison. In fact, while having the highest present-day luminosity of our combined sample, \A{XIII} has a stellar mass at $z=5$ that is comparable to some of the MW UFDs. Nonetheless, while the latter quenched rapidly, \A{XIII} halted star formation at a much later time ($t\lesssim 10$ Gyr). We discuss the SFH of this galaxy in greater detail in \S~\ref{sec:A13}.

The trends of Fig.~\ref{fig:SFHsample} can be quantified more clearly through a comparison of characteristic star formation timescales. This is a common practice in the field \citep[e.g.,][]{Weisz14,Skillman17, Weisz19, Sacchi21} which makes use of the quantity $\tau_{X}$; this is defined as the lookback time at which the galaxy reached $X\%$ of the total star stellar mass formed. For clarity, an illustration of these metrics is shown in Fig.~\ref{fig:SFHsample}. Measurements of select $\tau_{X}$ for our sample are provided in Tab.~\ref{tab:SFH}.

Figure~\ref{fig:T90} compares values of $\tau_{X}$ for the same MW and M31 galaxies shown in Fig.~\ref{fig:SFHsample}. It is again clear that, with the notable exception of \A{XIII}, all MW and M31 UFDs formed at least 50\% of their stellar mass by the end of cosmic reionization. Following previous studies \citep[e.g.,][]{Skillman17, Weisz19, Sacchi21} we can use $\tau_{90}$ as a tracer of the quenching epoch (Fig.~\hyperlink{T90}{5a, 5b}). This metric reveals a clear difference between MW and M31 UFDs.  MW UFDs quenched rapidly after reionization, whereas M31 UFDs sustained star formation until as late as $z\sim 1$. These extended SFHs can be appreciated in Fig~\hyperlink{T90}{5b}, which shows $\tau_{20-90}$, defined as the time elapsed between 20\% and 90\% of the total star formation. While the MW UFDs have $\tau_{20-90}$ of the order of 2 Gyr, M31's satellites tend to have significantly larger values ranging from 4 to 8 Gyr. 

In stark contrast to all other UFDs is \A{XIII}. It has a younger stellar population with $\tau_{50} = 10.6 \pm 0.6$ Gyr. Like other M31 UFDs in our sample, \A{XIII} shows signs of having some residual star formation ongoing for several Gyr, after the major event of star formation (which happened between 10 and 11 Gyr ago).

\subsection{Do M31's UFD Satellites Have Extended SFHs?} \label{sec:extended}

The paradigm established from the SFHs of MW UFDs is that they are ubiquitously ancient. Given that our M31 UFDs all show some degree of star formation post-reionization, in contrast to known MW UFDs, it is important that we assess the robustness of our findings.  We do this in three ways.  First, we have re-run our fits using different stellar population model assumptions (Appendix~\ref{App:models}), perturbations in distance and extinction (Appendix~\ref{App:Dgrid}), assessing SFH recovery as a function of photometric depth using other real data (Appendix~\ref{App:depth}), and employing a different CMD-fitting software (Appendix~\ref{App:Code}).  In short, the results of all these tests do not change the quantitative conclusions presented in this paper, with the exception of \A{XI}, which may be compatible with purely ancient star formation when different stellar models are used.   

Second, we consider the metric, $\tau_{X}$, by which we identify trends in the SFHs.  For quenching, $\tau_{90}$ has been historically used in the literature as it gets close to $\tau_{100}$, but mitigates known issues with low level of late star formation that can spuriously be introduced in the SFH by a variety of CMD contaminants, including photometric artifacts, foreground/background point sources, and blue straggler stars.

However, $\tau_{90}$ was introduced in the context of more massive galaxies with well-populated CMDs.  In the case of sparsely populated UFDs, even $\tau_{90}$ may be subject to spurious sources mimicking low-level star formation at late times.  Thus, in Figures~\hyperlink{T90}{5c} and~\hyperlink{T90}{5d} we adopt $\tau_{80}$ as a more conservative tracer of galaxy quenching. As expected, this choice results in significantly less extended SFHs compared to $\tau_{90}$. Nonetheless, indications of post-reionization star formation activity remain. This is most clearly the case for \A{XIII}.  It is also clear for \A{XXII}, which has a SFH that is inconsistent with quenching at $z=6$, at a $\sim2\sigma$ level. The remaining M31 UFDs are more consistent with the properties of the MW UFDs, although a moderate ($\gtrsim 1\sigma$) indication of post-reionization star formation is also present in \A{XX} and \A{XXVI}. Even with this conservative metric, we see similar hints of extended star formation in two MW UFDs (Ursa Major I and Hydra II), which, taken at face value, suggest that even the MW UFDs may not all be fully quenched by reionization.

\begin{figure*}

\plotone{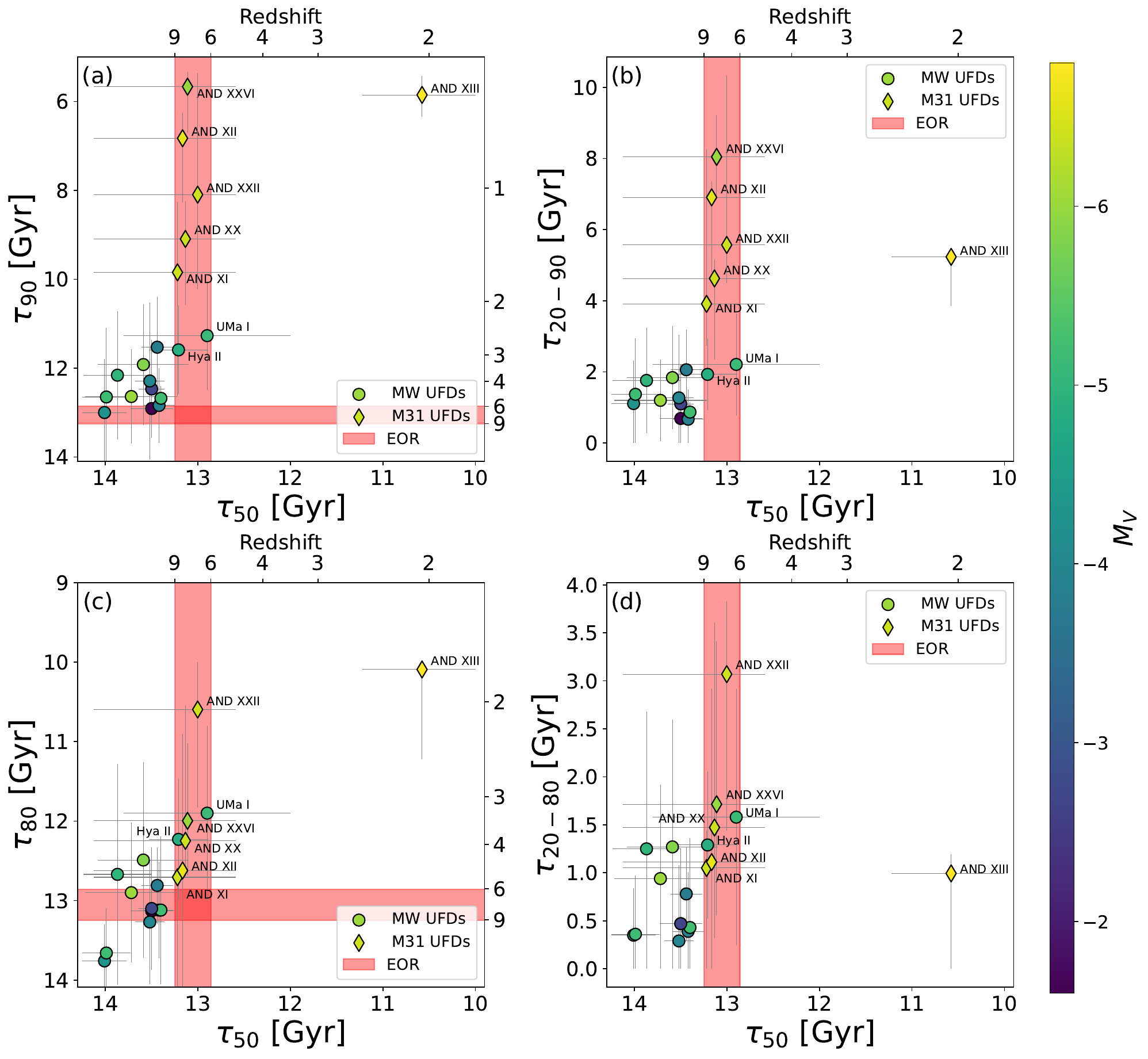}
\hypertarget{T90}{}

\caption{ Comparison of star formation timescales  (defined in \S~\ref{sec:MW}) for a sample of MW (circles) and M31 (diamonds) UFD satellites. Symbols are color-coded by the galaxy present-day luminosity. The red shaded region marks the approximate period of cosmic reionization \citep[][]{Robertson22}. a): time at 50\% of total star formation ($\tau_{50}$) vs time at 90\% of total star formation ($\tau_{90}$); b): $\tau_{50}$ vs time elapsed between 20\% and 90\% of total star formation ($\tau_{20}$-$\tau_{90}$); c): $\tau_{50}$ vs time at 80\% of total star formation ($\tau_{80}$); d): $\tau_{50}$ vs time elapsed between 20\% and 80\% of total star formation ($\tau_{20}$-$\tau_{80}$).}
\label{fig:T90}
\end{figure*}

Overall, the results of Fig.~\ref{fig:T90} indicate that, while the significance and duration of prolonged star formation depend somewhat on how we choose to define the signature of quenching, at least some of the M31 UFDs  have sustained some level of true star formation at $z\lesssim 5$. This conclusion is corroborated by additional studies of the M31 satellites. Generally extended SFHs, in excess of what is observed around the MW, have previously been obtained for other, more luminous M31 satellites from deep \hst\ imaging \citep{Weisz14b,Monelli16,Skillman17}. Similar tentative conclusions were previously reached for a larger M31 satellite sample (including some of the UFDs) based on SFHs from much shallower photometry \citep{Weisz14, Weisz19, McQuinn23}.  These SFHs necessarily have larger uncertainties in their SFHs, prohibiting the type of concrete conclusions we are able to draw in this paper. 

Third, we examine our results in light of the HB morphologies.  Recall that the HBs were excluded from the CMD fitting, meaning that they can provide an ancillary check on the reasonability of the MSTO-based SFHs.  The HBs of M31 satellites are known to be redder on average than their MW counterparts \citep[e.g.,][]{DaCosta96,DaCosta00,DaCosta02,Martin17}. This is also the case for our UFD sample. As noted in \S~\ref{sec:data}, \A{XIII} and \A{XXII}, which show the strongest indication of post-reionization star formation, also have predominantly red HB stars. Less extreme, but still significant, populations of red HB stars are also present in \A{XX} and \A{XXVI}, in accordance with their tentatively prolonged SFHs. The morphology of the HB has been previously shown to be a good tracer of the underlying SFH \citep[e.g.,][]{Salaris13,Savino18,Savino19} and, while lingering uncertainties remain on the absolute age calibration of HB stars, it is generally accepted that red HB morphologies are a sign of younger ages and/or higher metallicities compared to blue HBs.

Though the quality of spectrosopic metallicities available for M31 satellites greatly varies \citep[e.g.,][]{Collins13,Ho15,Kirby20}, what we do know so far is that they tend to follow the mass-metallicity relation known to exist for the MW satellites, and for thousands more galaxies in a large stellar mass range \citep[e.g.,][]{Tremonti04,Kirby13}. For the 6 UFDs in our sample, metallicity measurements have been obtained by \citet{Collins13}, although for a very limited number of spectroscopic members, and they suggest $\rm<[Fe/H]>$ values between $-2.3$ and $-1.9$, in good agreement with expectations from their stellar mass. In accordance with spectroscopy, our CMD fits report $\rm<[Fe/H]>$ values between $-2.3$ and $-1.6$, once the effect of $\alpha$-enhancement is taken into account. In an effort to obtain more secure measurements, members of our team have acquired deep Keck/DEIMOS spectroscopy of a larger red giant branch sample in the M31 UFDs, and preliminary analysis indicates they have low metallicities consistent previous measurements and with the mass-metallicity relation, which suggests the red HBs are more likely the result of younger ages and not higher metallicities (Sandford et al., in prep.).

\subsection{The Quenching of Local Group UFDs in the Context of Reionization} \label{sec:reionization}
Our analysis of M31's UFDs has interesting implications for our understanding of star formation and quenching in low-mass satellites. Studies consistently show that, below a certain halo mass threshold, the post-reionization UV background is able to prevent the accretion of new gas into the cold inter-stellar medium (ISM) reservoir \citep[e.g.,][]{Wheeler15,Dayal18,Hutter21}. It has also been proposed that the cold gas already present in the halo might be photoevaporated, thus halting star formation, although the effectiveness of this mechanism is debated \citep[e.g., ][]{Barkana99,Haiman01,Hoeft06,Busha10,Sawala10,Simpson13,Gutcke22}. The cosmic UV feedback has been therefore suggested to be the primary quenching mechanism for galaxies with stellar masses as high as $10^6 M_{\odot}$ \citep[e.g.,][]{Bovill09,Salvadori09}. 

Observationally, differences in SFHs between the long-known classical dwarfs \citep[e.g.,][]{Tolstoy98,Gallart99,Grebel04,Monelli10a,Monelli10b,deBoer14} and the more recently discovered MW UFDs \citep{Okamoto12, Brown14,Sacchi21,Simon21,Weisz23}, supported the suggestion of such evolutionary pathway for low-mass dwarfs and led to the conclusion that galaxies in the UFD class are uniformly ancient. The SFHs derived in this paper, for M31 UFDs, also show a general prevalence of ancient ($t<12.6$ Gyr) stars, and a subsequent lull of star formation activity. The timescale of this initial star formation burst, and its ubiquitous depression, are therefore compatible with the effect of cosmic reionization.

However, the properties of our sample, combined with literature measurements on the MW satellites, also suggest that the effect of reionization on UFDs is not homogeneous.  Variations appear to exist in how efficient the UV feedback is at permanently halting star formation. The prime example of this heterogeneity is \A{XIII} (which we discuss in detail in \S~\ref{sec:A13}). However,  evidence of inefficient quenching is also present in other galaxies of our sample. As discussed in \S~\ref{sec:extended}, the precise amount of post-reionization star formation  and the ultimate quenching epoch of our five ``ancient" UFDs (\A{XI}, \A{XII}, \A{XX}, \A{XXII}, and \A{XXVI}) is challenging to quantify (e.g., adopting $\tau_{90}$ vs $\tau_{80}$). Nonetheless, there are multiple lines of evidence that some of these galaxies have experienced subsequent episodes of star formation, building as much as 40\% of their total stellar mass over the course of a few Gyr after reionization.

Importantly, this finding has been expected by several theoretical studies. Multiple high-resolution simulations have indeed shown that the interplay among reionization, halo assembly history, and ISM properties, can result in a substantial fraction of star formation to happen at later times \citep[e.g.,][]{Hoeft06, Onorbe15, Fitts17,Jeon17,Munshi19,Rey20,Gutcke22}. In such a scenario, however, it still remains unclear what mechanism ultimately quenched our M31 UFDs, at a later redshift. Various suggestions include supernovae feedback \citep[e.g.,][]{Salvadori08,Sawala10,Gelli20,Gallart21} or environmental processes, such as ram pressure stripping, as the galaxies were accreted into the halo of M31 \citep[e.g.,][]{Mayer06,Romano15, Putman21}. Distinguishing between internal and environmental quenching mechanisms will likely require measuring the orbital history of our sample and pinning down their quenching redshift with greater accuracy.

It also remains to be determined what drives the systematic differences between the early SFH of the M31 and MW samples, as there are hard-to-quantify selection and physical effects at play. For instance, UFDs associated with different hosts might have spent their early life in different regions around the ``proto-Local Group". As reionization around the MW and M31 has been argued to be significantly patchy \citep[e.g.,][]{Ocvirk14,Ocvirk16,Aubert18,Sorce22}, current satellites of M31 might have experienced different reionization conditions (e.g., timing, UV background strength) compared to their MW counterparts. This scenario might be even more relevant in light of the extensive evidence that M31 has experienced more recent accretion events compared to the MW \citep[e.g.,][]{Belokurov18, Helmi18, McConnachie18, Helmi20}. These include massive galaxies (e.g. M33 and the progenitor of the Giant Stellar Stream), which are expected to have brought their own UFD satellites \citep[e.g.,][]{Chapman13, Patel18}. Satellites accreted at a later time might have spent their early life farther away from the strong ionizing sources around the proto-M31/MW. However, it is not clear if this difference in accretion history extends to the low-mass satellite population. In fact, it has been argued that, contrary to the lack of recent massive accretions, the MW has accreted UFDs at a relatively steady rate, up to recent times \citep[e.g.,][]{Wetzel15,Fillingham19,Rodriguez-Wimberly19,Santistevan23}, suggesting there may be  little difference in the ancient environments of UFDs.


On the other hand, our M31 sample also differs from the MW UFDs in terms of stellar mass. Due to detection biases \citep{Doliva-Dolinsky22}, M31's known UFDs are all relatively luminous, with present-day stellar masses of $\sim10^5 M_{\odot}$. At the time of reionization, these galaxies were already more massive than the MW sample (cf. Fig.~\ref{fig:SFHsample}) and might have resided in more massive and/or more concentrated dark matter halos\footnote{We highlight that this argument relies on the idea that the stellar mass in these galaxies roughly traces the depth of the gravitational potential. Given the expected scatter in the $M_{*}-M_{halo}$ relation at these low halo masses \citep[e.g.,][]{Garrison-Kimmel17}, this assumption is far from perfect. Halo masses, however, are notoriously difficult to determine in UFDs \citep[e.g., ][and references therein]{Simon19}. With only a handful of spectroscopic members being available in our M31 UFDs, dynamical masses within the half-light radius are only known within a factor of a few \citep{Collins13}. Extrapolations of these measurements to obtain a total halo mass are even more uncertain, as the stars only trace the inner $\sim 200$pc of the dark matter density profile.}. This would have increased their efficiency to cool the gas that was ionized by the UV background \citep[e.g.,][]{Jeon17,Benitez-Llambay20} and resume star formation at a later redshift. These galaxies might also have had a denser ISM, increasing the efficiency of self-shielding and allowing the retention of a cold gas reservoir \citep[e.g.,][]{Sawala10,Rahmati13,Rey20,Gutcke22}.


Galaxies in the $10^5 M_{\odot}$ stellar mass range are not common around the MW, which could explain the lack of extended SFHs. In fact, the study of galaxies at the massive end of the UFD spectrum was one of the main goals behind our M31 satellite survey. While providing an elegant explanation for the differences between MW and M31's UFDs, this scenario also has counterexamples. While not common, a few luminous UFDs exist around the MW (e.g., Bootes I or Hercules dSph). If galaxy mass was the dominant factor behind the efficiency of UFD quenching, we would expect these galaxies also to have extended SFHs but their measured ages are ancient \citep{Brown14}. Conversely, the SFH of \A{XIII} implies that this galaxy was at most $ 4\cdot10^4 M_{\odot}$ at the time of reionizaion, and had about similar stellar mass to some MW UFD satellites, such as Coma Berenices or Canes Venatici II. Yet, while the latter have quenched rapidly after $z=6$, \A{XIII} has kept forming stars. However, the SFH of \A{XIII} is sufficiently anomalous that it deserves a separate discussion, as it might point to a completely different evolutionary pathway.

\subsection{The Case of \A{XIII}: the First Observed Episode of Reignition in an Ultra-Faint Dwarf} \label{sec:A13}

Unlike the five other galaxies in our sample, \A{XIII} experienced little star formation prior to $z\sim3$. At this time, it experienced a strong star formation episode and formed most of its stellar mass. This type of SFH is the first of its kind observed in such a low-mass system.  Galaxies such as Leo~T, \A{XVI}, and \A{XIX} have SFHs that exhibit similar pauses \citep[e.g.,][]{Irwin07,deJong08,Clementini12,Weisz14b,Monelli16,Skillman17,Collins22b}, but they are more luminous systems ($M_V\leq-7.5$). This general type of behavior (i.e., ``reignition'' or late ignition) has been seen in several cosmological simulations of dwarf galaxy populations by different groups \citep[e.g.,][]{Maccio17,Fitts17, Digby19, Jeon19, Garrison-Kimmel19,Wright19, Benitez-Llambay21, Applebaum21}, but typically in systems that are more massive ($M_*\gtrsim5\times 10^5 M_{\odot}$) than \A{XIII}. 

Within the assumption that the ability to fuel a cold gas reservoir in low-mass halos is regulated by the competition of radiative cooling and photoheating from the cosmic UV background, one possible scenario to explain the peculiar SFH of \A{XIII} relies on an unusual mass assembly history. The expectation from $\Lambda$CDM is that, at a given redshift, efficient gas cooling can only proceed in halos above a critical mass $M_{cr}(z)$ \citep[e.g.][]{Gnedin00,Okamoto08,Fitts17,Benitez-Llambay20, Pereira-Wilson23}. In the pre-reionization era, the value of $M_{cr}$ is dictated by the atomic hydrogen cooling limit, while the post-reionization threshold is set by the properties of the inter-galactic medium  and the strength of the UV background. If the halo that would eventually host \A{XIII} had an unusually slow accretion rate at early times, so that $M_h(z) \leq M_{cr}(z)$, it might have reached the reionization epoch having formed only a modest amount of stellar mass. At this stage, the properties of \A{XIII} could have been similar to those of some of the quenched UFDs around the MW. After reionization, the mass of \A{XIII}'s halo would have remained below $M_{cr}$ until $z\simeq3$, after which it experienced a substantial increase in mass growth rate. The deeper gravitational potential would have then allowed \A{XIII} to reconstitute a reservoir of cold ISM and eventually ignite prominent star formation. Adopting the classification scheme of \citet{Gallart15}, \A{XIII} would be therefore a very-low mass example of a ``slow" dwarf (whereas the remaining five UFDs of our sample would be ``fast" dwarfs).

An alternative scenario is that proposed by \citet{Wright19}, in which star formation is triggered by the interaction with a dense gas pocket (either in a filament or in the circumgalactic medium of a more massive galaxy). Such interaction would have compressed the hot gas around \A{XIII}, allowing cooling and enabling star formation reignition, without the need of strong dark matter accretion.

A third hypothesis is that \A{XIII} underwent a major merger with another UFD of comparable mass. This scenario combines both mechanisms discussed above. A merger would rapidly increase the dark matter halo mass, resulting in a deeper potential, and also result in compression of any residual gas present in the two system. This event could therefore be followed by a vigorous reprise of star formation. The occurrence rate and the products of major mergers in UFDs have not been explored extensively. Simulations of more massive classical dwarfs suggest that major mergers can occur and have the ability to reignite star formation \citep[e.g.][]{Benitez-Llambay16}. Furthermore, while dwarf-dwarf mergers are expected to be rare in the present Universe, their occurrence rate has been shown to increase at higher redshift \citep[e.g.,][]{Deason14}, potentially being a viable channel for the reignition of star formation in \A{XIII}. 

Finally, we note that the unusual SFH of \A{XIII} is compatible with the expected properties of tidal debris arising from major galaxy interactions. However, both the observations of suspected tidal dwarfs and simulations of these objects \citep[e.g., ][]{Duc14,Ploeckinger15, Ploeckinger18,Gray23}, suggest that tidal dwarfs manifest as high-metallicity, gas-rich systems. The low-metallicity, gas-poor nature of \A{XIII}, therefore, appears to rule out a tidal origin.

Irrespective of the mechanism responsible for the late star formation ignition, it is interesting to compare \A{XIII}'s SFH with expectations from galaxy formation models. As mentioned already, instances of late-forming low-mass ($10^{5-6}M_{\odot}$) dwarf galaxies appear in different cosmological simulations \citep[e.g.,][]{Maccio17,Garrison-Kimmel19,Benitez-Llambay21,Applebaum21}. The properties of \A{XIII}, therefore, are not necessarily in tension with $\Lambda$ cold dark matter predictions. Is it worth noting, however, that such galaxies are a relatively uncommon occurrence in those models, with most simulated objects characterized by predominantly ancient SFHs. The expected frequency of late star formation in low-mass dwarfs has not been characterized in detail and is likely to vary among different simulations. For instance, \citet{Benitez-Llambay21} report that $\sim8\%$ of their simulated dwarfs with $10^{9.5}\lesssim M_{200}/M_{\odot}\lesssim 10^{10}$ begin formation after $z\sim3$, while \citet{Wright19} report that $\sim20\%$ of their dwarfs, in a comparable mass interval, experience reignition at low redshift.

Likewise, the discovery of \A{XIII}'s late SFH, and of other more massive dwarfs with similar behavior (e.g., \A{XVI}), is not yet sufficient for a statistically robust estimate of the abundance of such objects. Nevertheless, as we discover more low-mass dwarfs in the Local Group and its vicinity, we will have a clearer picture of how common low-mass late-forming dwarfs are in the local Universe and how this compares to the abundance expected from galaxy formation models.


 \subsection{Future Prospects}
The SFHs derived in this paper expand our observational picture of galaxy formation in low-mass halos, by studying the evolution of UFDs in a different satellite system and beginning to explore the ``transition" regime between low-mass reionization fossils and more massive classical dwarfs \citep[e.g,][]{Ricotti05}. Further work, however, is required to address open questions, such as delineating the relationship between stellar and halo mass growth for UFDs in the early Universe, bracketing the critical mass range at which the UV background is not as effective as quenching faint galaxies, and identifying the drivers of star formation suppression in the UFDs that resume star formation after reionization.

We anticipate that important observational advancements will be enabled by the synergy of large-area deep surveys \citep[e.g., those of the  Vera C. Rubin Observatory and Nancy Grace Roman Space Telescope, ][]{Spergel15,Zeljko19}, which will increase the census of known UFDs over a range of stellar masses \citep[e.g.,][]{Mutlu-Pakdil21}, with high angular resolution follow-ups, through \hst\ and \textit{JWST} and \textit{Roman}, which will allow in-depth studies of these objects (i.e., SFHs and orbital histories).

It remains unclear how much of the SFH diversity revealed by Figs.~\ref{fig:SFHsample} and \ref{fig:T90} is due to differences in stellar mass and 
how much is caused by the different environmental histories of M31 and MW satellites. A first development on this front will be the discovery and characterization of the faint ($M_* \lesssim 10^5 M_{\odot}$) UFD population around M31. This will enable a direct comparison between MW and M31 UFD satellites of similar mass and unveil whether systematic differences exist in the SFHs of the two UFD systems.

Second, since our Treasury program has established the first-epoch imaging of the six UFDs of our paper, future observations will be able to obtain precise proper motions for these systems, as is already starting to be demonstrated by \citet{Warfield23}. In turn, this will enable the reconstruction of orbital histories and the estimation of infall times, as has been done for the MW satellites \citep[e.g.,][]{Gaia18,Kallivayalil18,Fritz18,Fillingham19,Patel20,Battaglia22}. This will shed light on the role that M31's low-mass satellite accretion history had in shaping the evolution of its UFDs. Orbital histories will also enable the identification of UFDs associated with some of the most massive M31 satellites, as demonstrated for the MW and the Large Magellanic Cloud \citep[e.g.,][]{Kallivayalil18,Erkal19,Pardy20,Patel20}. In this light, we note that \A{XXII}, which has strong indication of an extended SFH, has been tentatively proposed as a satellite of M33 \citep[e.g.,][]{Martin09,Tollerud12,Chapman13}. Proper motions will allow to confirm or refute this association, and assess whether it has experienced a significantly different environmental history compared to the remaining UFDs.

Third, future surveys will discover UFDs with $10^5 M_{\odot} \lesssim M_* \lesssim 10^6 M_{\odot}$ in isolated environments outside the Local Group \citep[e.g.,][]{Mutlu-Pakdil21,Sand22,Qu22}. The SFHs of these systems will prove critical to establish whether the star formation timescales and quenching (or lack thereof) in these systems are fundamentally different from those of the satellite systems of massive hosts. This will help disentangle the role of environmental effects from other more universal drivers of star formation suppression, especially in regard to their post-reionization evolution.

Finally, the discovery of a peculiar UFD such as \A{XIII}, and differences between M31 UFDs and the most massive MW UFDs, suggest that a surprising diversity of SFHs might be present in UFDs in this ``transition" mass range ($10^{4.5}\lesssim \frac{M_{*}}{M_{\odot}}(z=0) \lesssim 10^{5.5}$). A better depiction of how galaxy formation proceeds in this mass range will require a larger sample of massive UFDs. Thanks to future large area surveys, we expect that such objects will probably be discovered in the outer halo of M31, in the periphery of the Local Group and in its immediate vicinity.

\vspace{4mm}
\section{Conclusions}
Using deep \hst\ imaging that reaches the oMSTO, we derived lifetime SFHs for a sample of six UFD galaxies associated with the satellite system of M31. These are the first oMSTO-based SFHs for UFD galaxies outside of the MW halo. The main takeaways of our analysis are:

\vspace{2mm}
i) Five of the UFDs in our sample formed at least 50\% of their stellar mass by $z=5$ (12.6 Gyr ago), after which star formation activity drastically reduced in intensity. This is similar to other known UFDs and is compatible with the predicted effect of cosmic reionization. However, we find that star formation is not entirely quenched in our galaxies and 10-40\% of their stellar mass formed at later times.

\vspace{2mm}
ii) One UFD in our sample (\A{XIII}) has a remarkable SFH, having formed only $\sim 10\%$ of its stellar mass by $z=5$ and having experienced a delayed period of strong star formation at $2<z<3$. This UFD is the first of its kind and suggests that star formation reignition, previously observed in other more massive galaxies, can also occur at lower stellar masses.

\vspace{2mm}
iii) Combining SFHs for MW and M31 UFDs, we conclude that the effect of reionization is not homogeneous among UFDs, with more massive galaxies being able to sustain low levels of prolonged star formation compared to low-mass UFDs, which are quenched more rapidly. However, due to selection biases, the environmental dependence of this trend cannot be ruled out.

\vspace{2mm}
iv) Future observational efforts, including large-area surveys, and dedicated follow-ups with \hst\ and \textit{JWST}, will provide an extended observational baseline to address some of the remaining open questions in UFD evolution. These include a better understanding of the role of cosmic environment, establishing the degree of SFH diversity in galaxies at the massive end of the UFD spectrum, and bracketing the stellar/halo mass range where the transition between ancient reionization fossils and classical dwarf galaxies occurs.


\begin{acknowledgments}
The authors thank the anonymous referee for a constructive report. AS thanks E. Sacchi for graciously sharing the star formation histories of seven Milky Way satellites. Support for this work was provided by NASA through grants GO-13768, GO-15476, GO-15902, AR-16159, and GO-16273 from the Space Telescope Science Institute, which is operated
by AURA, Inc., under NASA contract NAS5-26555. AW acknowledges support from: NSF via CAREER award AST-2045928 and grant AST-2107772; NASA ATP grant 80NSSC20K0513; \hst\ grants AR-15809, GO-15902, GO-16273 from STScI. MCC acknowledges support though NSF grant AST-1815475. MBK acknowledges support from NSF CAREER award AST-1752913, NSF grants AST-1910346 and AST-2108962, NASA grant NNX17AG29G, and HST-AR-15006, HST-AR-15809, HST-GO-15658, HST-GO-15901, HST-GO-15902, HST-AR-16159, and HST-GO-16226 from STScI. ENK acknowledges support from NSF CAREER grant AST-2233781. This research has made use of NASA’s Astrophysics Data System Bibliographic Services.
\end{acknowledgments}

\vspace{5mm}
\facilities{ \hst\ (ACS). All the data presented in this paper were obtained from the Mikulski Archive for Space Telescopes (MAST) at the Space Telescope Science Institute. The specific observations analyzed can be accessed via \dataset[DOI: 10.17909/ftep-8k64]{https://doi.org/10.17909/ftep-8k64}.
}

\software{ This research made use of routines and modules from the following software packages: \texttt{Astropy} \citep{Astropy}, \texttt{DOLPHOT} \citep{Dolphin16}, \texttt{IPython} \citep{IPython},
\texttt{MATCH} \citep{Dolphin02}, \texttt{Matplotlib} \citep{Matplotlib}, \texttt{NumPy} \citep{Numpy}, \texttt{Pandas} \citep{Pandas}, and \texttt{SciPy} \citep{Scipy}.}


\bibliography{M31_UFD}{}
\bibliographystyle{aasjournal}

\appendix

\begin{figure*}
\plotone{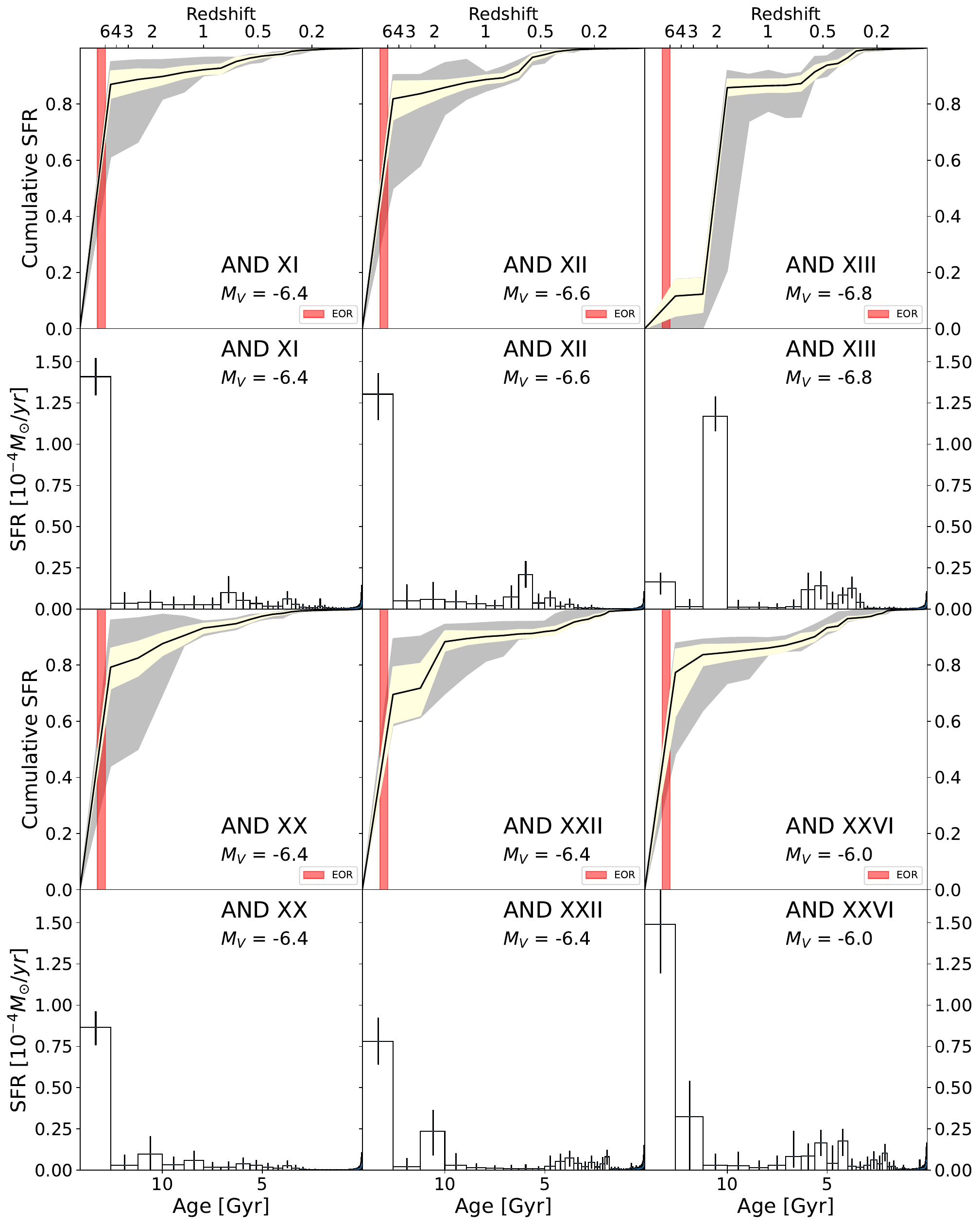}

\caption{Same as Fig.~\ref{fig:SFHs}, but including both cumulative (top) and differential star formation histories.}

\label{fig:SFR}
\end{figure*}

\section{Instantaneous star formation rates}
\label{App:SFH}

Here we provide the instantaneous star formation rates, also referred to as differential SFHs, derived from our fiducial solutions. These are illustrated in Fig.~\ref{fig:SFR}, alongside the corresponding cumulative SFHs. Differential SFHs offer an intuitive way to visualize the star formation activity of the galaxy at any point in time. However, it is important to excercise caution when interpreting the uncertainties associated with these measurements. The uncertainties in the histograms, in fact, come from the variance of the star formation rate within each time bin. On the other hand, due to the details of the CMD fitting method \citep[see, e.g.,][]{Dolphin02} significant covariance exists among adjacent star formation bins. This constrains the range of plausible solutions that accurately fit the CMD.

For instance, in Fig.~\ref{fig:SFR}, most of the late star formation bins have error bars compatible with zero star formation (see, e.g., \A{XX}). However, the star formation rate \textit{cannot} be zero simultaneously in all bins, as that would significantly impact the star counts within the CMD. To address this limitation, cumulative SFHs are conventionally preferred, because they are constructed to incorporate the covariance between star formation rates. In the case of \A{XX}, the statistical uncertainties on the cumulative SFH (yellow shading), unequivocally reject a complete absence of star formation after the oldest time bin.

\begin{figure*}
\plotone{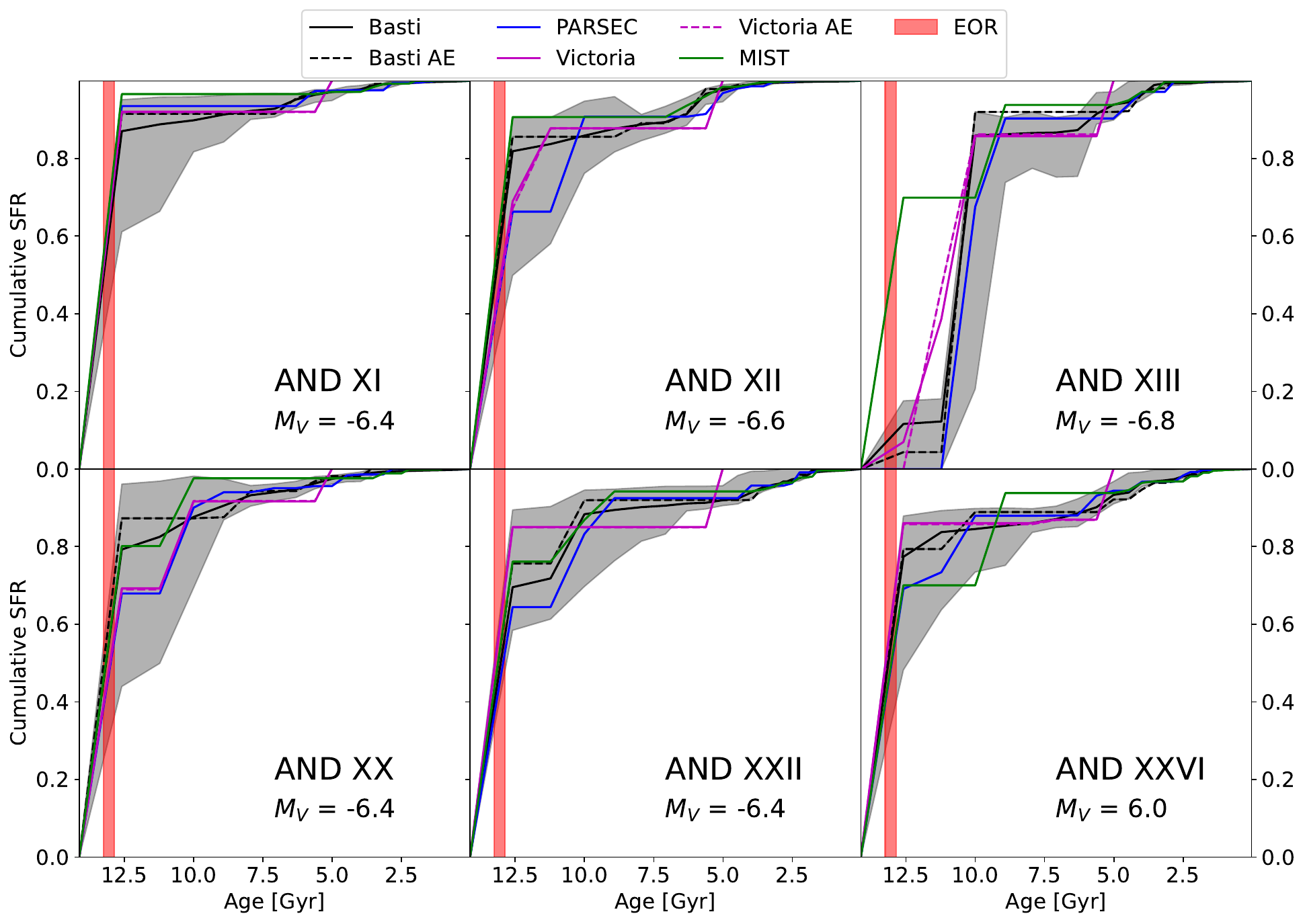}

\caption{Alternative SFH solutions for our sample, obtained with the PARSEC (blue), MIST (green) and Victoria-Regina (magenta) stellar libraries, along with our fiducial solution that uses BaSTI (black). Solid lines indicate scaled-Solar models while dashed lines indicate $\alpha$-enhanced models. The shaded grey region outlines the systematic uncertainties estimated through the method of \citet{Dolphin12}. The red shaded region marks the epoch or reionization.  In most cases these are only modest differences in the SFHs as a function of stellar model, confirming the robustness of our results.
}
\label{fig:Models}
\end{figure*}

\section{The effect of stellar population models}
\label{App:models}
One of the primary sources of systematic uncertainties in the CMD modeling of external galaxies lies in the choice of stellar evolution models \citep[e.g.,][]{Gallart05}. In the main text, we have estimated the magnitude of these effects by using the methodology of \citet{Dolphin12}, which generates a large set of mock stellar models by applying systematic shifts in luminosity and effective temperature to the adopted stellar evolutionary tracks. Here, we provide a different estimation of the stellar evolution systematics, by re-running our analysis with a range of different stellar evolution models. In addition to our fiducial stellar model grid \citep[BaSTI, ][]{Hidalgo18}, we have modeled the CMDs of our targets using the PARSEC \citep{Bressan12, Chen15}, MIST \citep{Dotter16, Choi16} and Victoria-Regina \citep{Vandenberg14} stellar models. Furthermore, while our fiducial analysis has been done using a scaled-Solar abundance mixture, here we assess the effect of the assumed alpha-enhancement by including the $[\alpha/Fe] = 0.4$ models of the BaSTI \citep{Pietrinferni21} and Victoria-Regina grids.

\begin{table}[]
    \centering
    \begin{tabular}{lcccccc}
        \toprule
        Library & \A{XI} & \A{XII} & \A{XIII} & \A{XX} & \A{XXII} & \A{XXVI} \\
        \toprule
        &&&$\tau_{50}$&&\\
        &Gyr&Gyr&Gyr&Gyr&Gyr&Gyr\\
        \toprule
        Basti SS & 13.2& 13.2& 10.6& 13.1& 13.0&13.1\\
        Basti AE & 13.3& 13.2& 10.6& 13.2& 13.1&13.1\\
        PARSEC & 13.3& 13.0& 10.3& 13.0& 12.9& 13.0\\
        MIST & 13.3& 13.3& 13.0& 13.1& 13.1&13.0\\
        Victoria SS & 13.3& 13.0& 10.9& 13.0& 13.2&13.2\\
        Victoria AE & 13.3& 13.0& 11.1& 13.0& 13.2&13.2\\
        \toprule
        &&&$\tau_{80}$&&\\
        &Gyr&Gyr&Gyr&Gyr&Gyr&Gyr\\
        \toprule
  
        Basti SS &12.7& 12.6& 10.1& 12.3& 10.6& 12.0\\
        Basti AE & 12.8& 12.7& 10.2& 12.7& 10.9&11.1\\
        PARSEC & 12.8& 10.5& 9.4& 10.5& 10.2& 10.6\\
        MIST & 12.8& 12.8& 9.5& 12.6& 10.8&9.5\\
        Victoria SS & 12.8& 11.8& 10.1& 10.6& 12.7&12.7\\
        Victoria AE & 12.8& 11.7& 10.2& 10.6& 12.7&12.7\\
                \toprule
        &&&$\tau_{90}$&&\\
        &Gyr&Gyr&Gyr&Gyr&Gyr&Gyr\\
        \toprule
        Basti SS & 9.8& 6.8& 5.9& 9.1& 8.1&5.7\\
        Basti AE & 12.6& 6.8& 10.0& 8.5& 10.1&5.4\\
        PARSEC & 12.6& 10.0& 8.9& 10.0& 9.2&6.0\\
        MIST & 12.7& 12.6& 9.1& 10.5& 9.5&9.1\\
        Victoria SS & 12.6& 5.5& 5.4& 10.1& 5.4&5.5\\
        Victoria AE & 12.6& 5.5& 5.4& 10.1& 5.4&5.5\\
        \toprule
    \end{tabular}
    \caption{{Values of $\tau_{50}$, $\tau_{80}$, and $\tau_{90}$ obtained, for the 6 galaxies of our sample, by using different stellar evolution libraries.}}
    \label{tab:stellar_models}
\end{table}

\begin{figure*}
\plotone{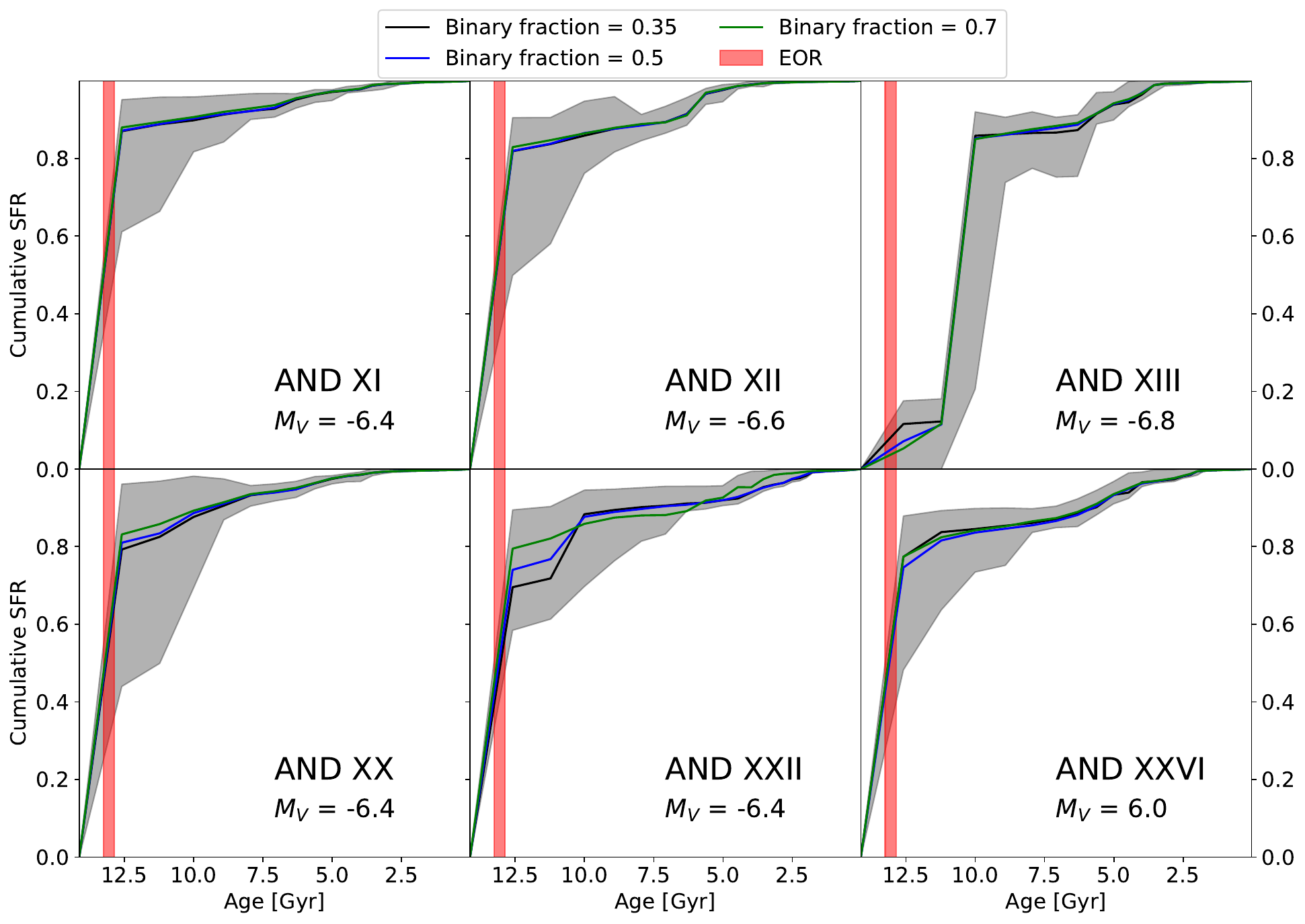}

\caption{Alternative SFH solutions for our sample, obtained with an unresolved binary fraction of 0.5 (blue) and 0.7 (green), along with our fiducial solution that uses a binary fraction of 0.35 (black). The shaded grey region outlines the systematic uncertainties estimated through the method of \citet{Dolphin12}. The red shaded region marks the epoch or reionization.  The choice of binary fraction has a negligible effect on our solutions.} 

\label{fig:BF}
\end{figure*}

 \begin{figure*}
\plotone{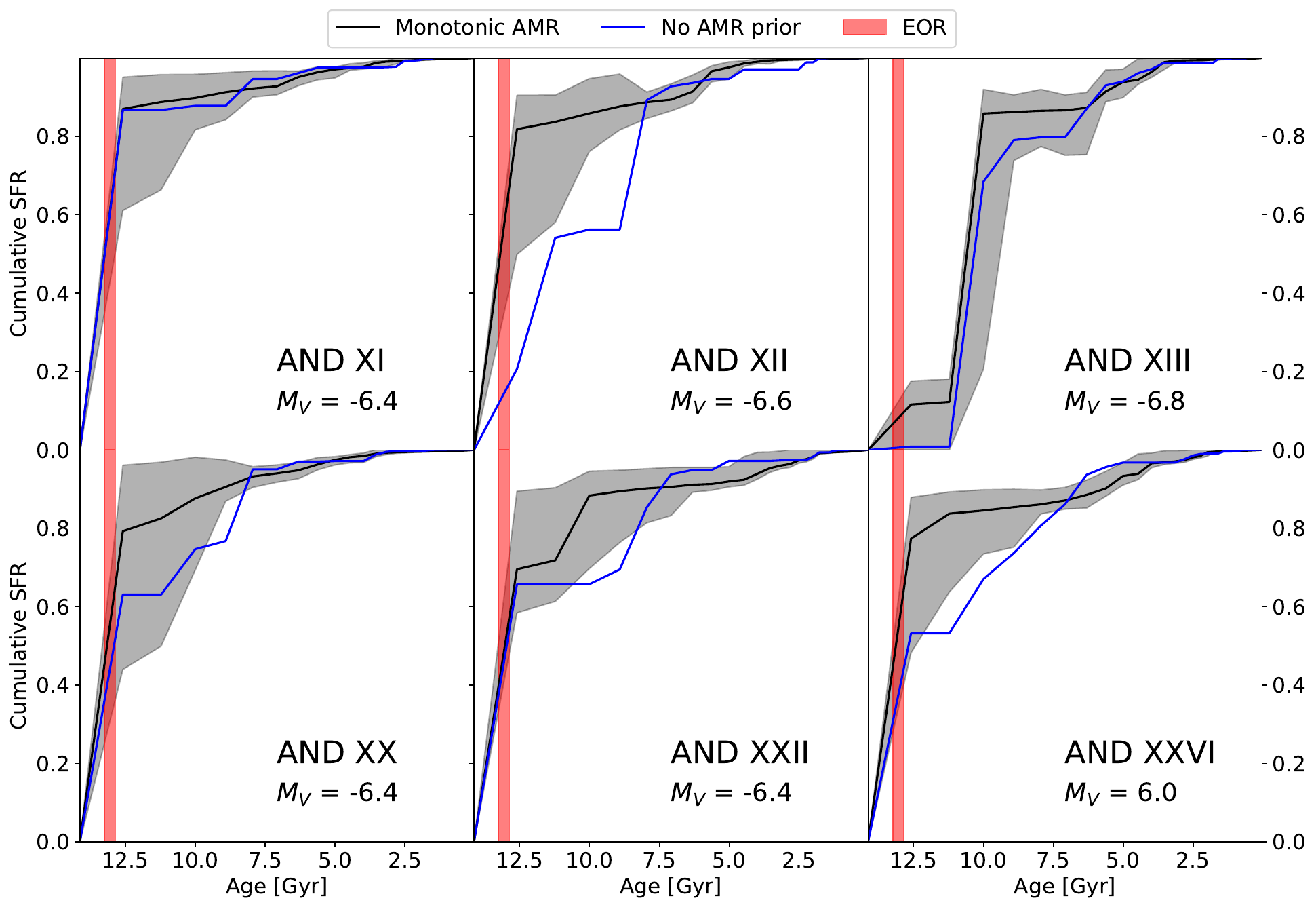}

\caption{Alternative SFH solutions for our sample, obtained by removing our constraint on the AMR shape (blue), along with our fiducial solution that assumes a monotonically increasing AMR (black). The shaded grey region outlines the systematic uncertainties estimated through the method of \citet{Dolphin12}. The red shaded region marks the epoch or reionization. Changing our assumptions on the shape of the AMR does not remove the presence of late star formation in our targets.}

\label{fig:AMR}
\end{figure*}

The resulting SFHs are shown in Fig.~\ref{fig:Models}. First, we note that the adopted $\alpha$-enhancement has little effect on the shape of the SFH. The scaled-Solar and $\alpha$-enhanced solutions are virtually indistinguishable in most cases. In this regard, the major difference that the $\alpha$ enhancement makes is to introduce a zero-point shift in the [Fe/H] scale of our solutions.  This is a well-known effect from stellar evolution theory \citep[e.g., ][]{Salaris93}. Second, the range spanned by the different SFH solutions is well-captured by the systematic confidence interval from the \citet{Dolphin12}. The latter uncertainty interval tends to be slightly larger compared with the test presented here. But, as discussed in \citet{Dolphin12}, the systematics are designed to be conservative given that there are only a limited set of stellar libraries available for use; they are unlikely to include the full range of reasonable stellar evolution uncertainties. Finally, the qualitative conclusions drawn from our SFHs are broadly unaffected by the choice of stellar models. With the exception of \A{XIII}, all our galaxies have a predominantly ancient population, regardless of stellar model choice. However, all stellar libraries suggests that these galaxies experienced significant star formation (10-40\% of total stellar mass) after reionization, with the only exception of \A{XI}, which appears purely ancient.

For \A{XIII}, three out of the four stellar libraries used here confirm that the galaxy had little to no star formation at ancient times. The BaSTI and PARSEC models date the start of significant star formation at around $z\simeq 3$, while the Victoria-Regina models put it around $z\simeq 4$. Only the MIST stellar library allows for the presence of ancient stars in this galaxy and, even then, at least 30\% of the stellar mass must have formed at a later time. It must be noted that, while the MIST library estimates an older age for \A{XIII}, compared to the other models, the same happens for the other 5 UFDs. For these galaxies, in the absence of our cosmological limit of 14 Gyr on the grid edge, the best fit SFH from MIST would peak at ages much older than the age of the Universe. This means that, even with the MIST models, \A{XIII} remains a younger galaxy compared to the rest of the UFD sample.

The values of $\tau_{50}$, $\tau_{80}$, and $\tau_{90}$ measured from the different solutions of Fig.~\ref{fig:Models} are listed in Tab.~\ref{tab:stellar_models}. The measured values of $\tau_{50}$ are in very good agreement, with an average standard deviation of $\sim 200$ Myr across the different stellar evolution libraries. The values of $\tau_{80}$ and $\tau_{90}$ have higher standard deviations of $\sim 700$ Myr and $\sim 1.6$ Gyr, respectively. These characteristic spreads among the models is comparable to the uncertainties listed in Tab.~\ref{tab:SFH}. Overall, the result of these tests is that, while the choice of stellar models can affect the details of the measured SFHs by a modest degree, the key SFH features identified in the paper are robust, regardless of stellar library adopted.

As noted in \S~\ref{sec:model}, we have also tested the effect of adopting different binary fractions, by rerunning our fits with a binary fraction of 0.5 and 0.7. These solutions are illustrated in Fig.~\ref{fig:BF}, along with our fiducial SFHs, which use a binary fraction of 0.35. The results of this test is that the choice of binary fraction has an almost negligible effect on the recovered star formation history, with only \A{XXII} showing a modest difference in the amount of star formation in the oldest bin.

An additional test on the stellar population models regards the prior on the age-metallicity relation (AMR). Our fiducial SFHs are based on the assumption that the AMR monotonically increases with time. This is a common practice \cite[e.g.,][]{Weisz11,Weisz14,McQuinn15} that results in physically reasonable solutions even at low S/N on the oMSTO. However, given the results of this paper, especially on \A{XIII}, a monotonically increasing AMR might not necessarily be an appropriate assumption. For instance, the late star formation we measure in some of our targets could have been triggered by the accretion of low-metallicity gas. We have therefore calculated a set of alternative SFH solutions, in which the AMR is left free to assume any form.

The result of this test are shown in Fig.~\ref{fig:AMR}. Removing the constraints on the AMR has, for the most part, a modest effect. Most notably, \A{XIII} remains a  purely ``young" galaxy, while \A{XI} maintains its predominantly ancient SFH. The effect on the other targets is more pronounced, especially in \A{XII}. However, changing our assumptions on the AMR does not impact the main conclusion of the paper, i.e., the presence of post-reionization star formation in some UFDs of our sample. On the contrary, the alternative SFHs of Fig.~\ref{fig:AMR} have an higher amount of late star formation, compared to our fiducial solutions. However, we advise strong caution in interpreting these results. By removing the requirement of a monotonic AMR, we have dramatically increased the degrees of freedom in our solutions, which can result in an unphysical metallicity evolution. In the case of Fig.~\ref{fig:AMR}, the new solutions have strongly fluctuating AMRs, with values of [M/H] approaching Solar values and varying by more than 1~dex between adjacent time bins. Nevertheless, a safe conclusion from this experiment is that our assumptions on the AMR do not impact the presence of late star formation in our sample, especially with respect to the case of \A{XIII}.

\begin{figure*}
\plotone{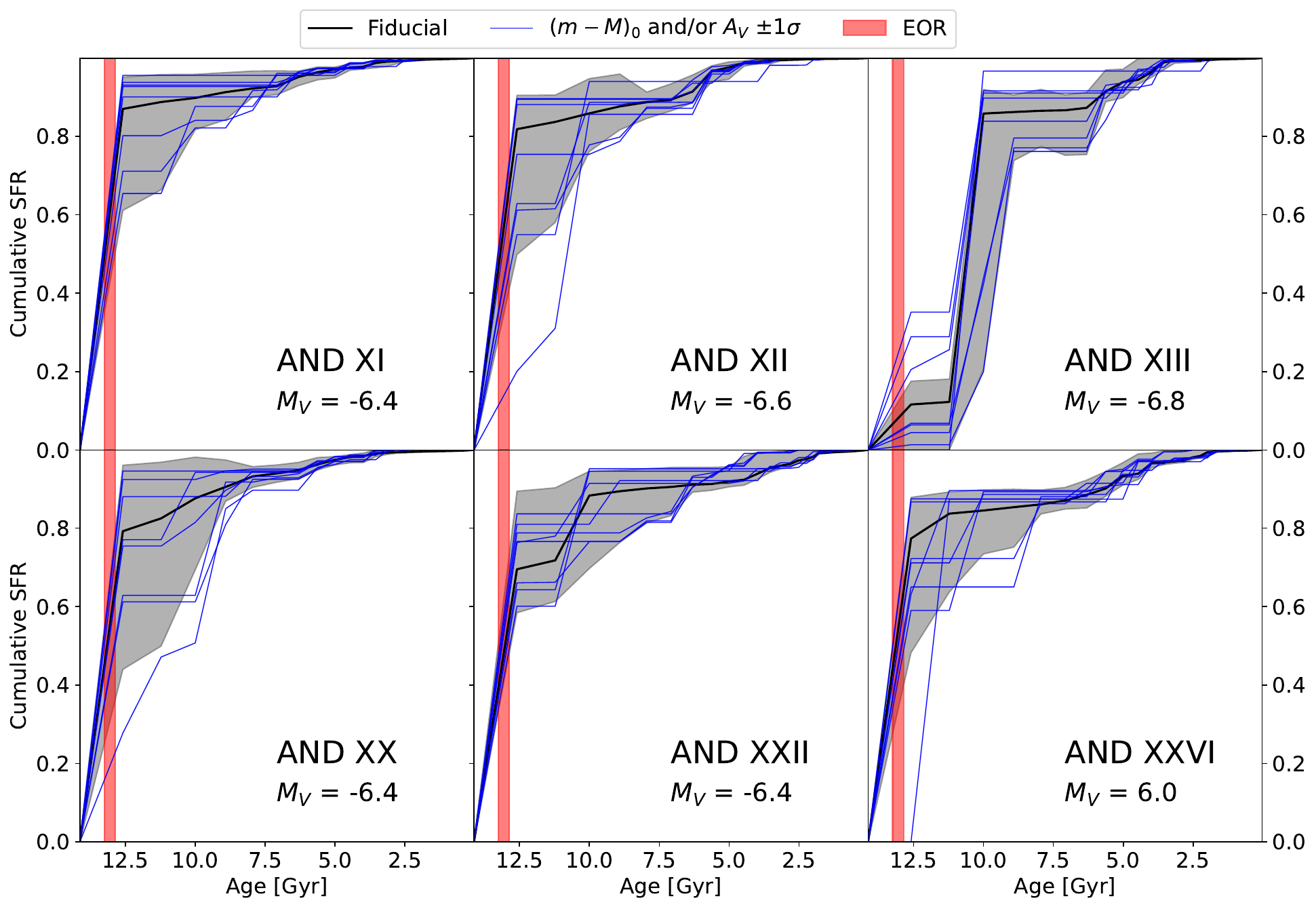}

\caption{Cumulative SFHs for our sample of UFD obtained by systematically varying distance and foreground extinction by $\pm 1 \sigma$ (blue lines). The black line shows the fiducial solution obtained in \S~\ref{sec:model}. The shaded grey region outlines the systematic uncertainties estimated through the method of \citet{Dolphin12}. The red shaded region marks the epoch of reionization.}
\label{fig:Dgrid}
\end{figure*}

\section{The effect of distance and reddening uncertainties}
\label{App:Dgrid}
Two assumptions made in SFH measurements are the distance of the observed galaxy and the amount of foreground dust extinction. These will affect the magnitude and color of the observed stars and, potentially the SFH inferred by CMD fitting. To assess the effect of these parameters on our SFH measurements we have repeated the analysis varying the assumed distance and reddening values, in accordance with the reported uncertainties in the literature. We have adopted distance uncertainties from \citet{Savino22}. For the 6 dwarf galaxies of our sample, these are of the order of 0.07-0.08 mag. Reddening uncertainties were taken from the dust maps of \citet{Green19} and amount to 0.03 mag in E(B-V). We have then calculated a set of alternative SFHs, in a $3\times3$ grid, by applying [$-1\sigma$, 0, $+1\sigma$] shifts in distance and [$-1\sigma$, 0, $+1\sigma$] shifts in reddening to our CMD models (the central point of this grid is the trivial case where no shifts are applied and corresponds to our fiducial SFH solution.).


The results of these tests are illustrated in Fig.~\ref{fig:Dgrid} as blue lines and compared with the fiducial SFH used in the paper (black). The effect of distance and reddening uncertainties is comparable in size with other sources of systematic uncertainties included in our error budget (grey region in Fig.~\ref{fig:Dgrid}). They modestly change the details of the SFHs but overall preserve the results presented in the paper. Most of the solutions for \A{XI}, \A{XII}, \A{XX}, \A{XXII} and \A{XXVI} have a main burst of star formation at the oldest ages, followed by more recent star formation accounting for 10 to 40\% of the total stellar mass. Conversely, \A{XIII} maintains a significantly delayed star formation compared to the other dwarfs.

We note from Fig.~\ref{fig:Dgrid} that certain combinations of distance and reddening can result in substantial changes in the inferred SFH. For the cases of \A{XII}, \A{XX} and \A{XXVI} this manifests as a significantly younger solution compared to the other set-ups. These solutions correspond to the extreme points of our parameter space, where both distance and reddening are shifted by 1 $\sigma$. Under the assumption of Gaussian uncertainties, the chance of this happening for a single galaxy is  $\sim10$\% and much lower for the entire set of galaxies.

\begin{figure*}
\plottwo{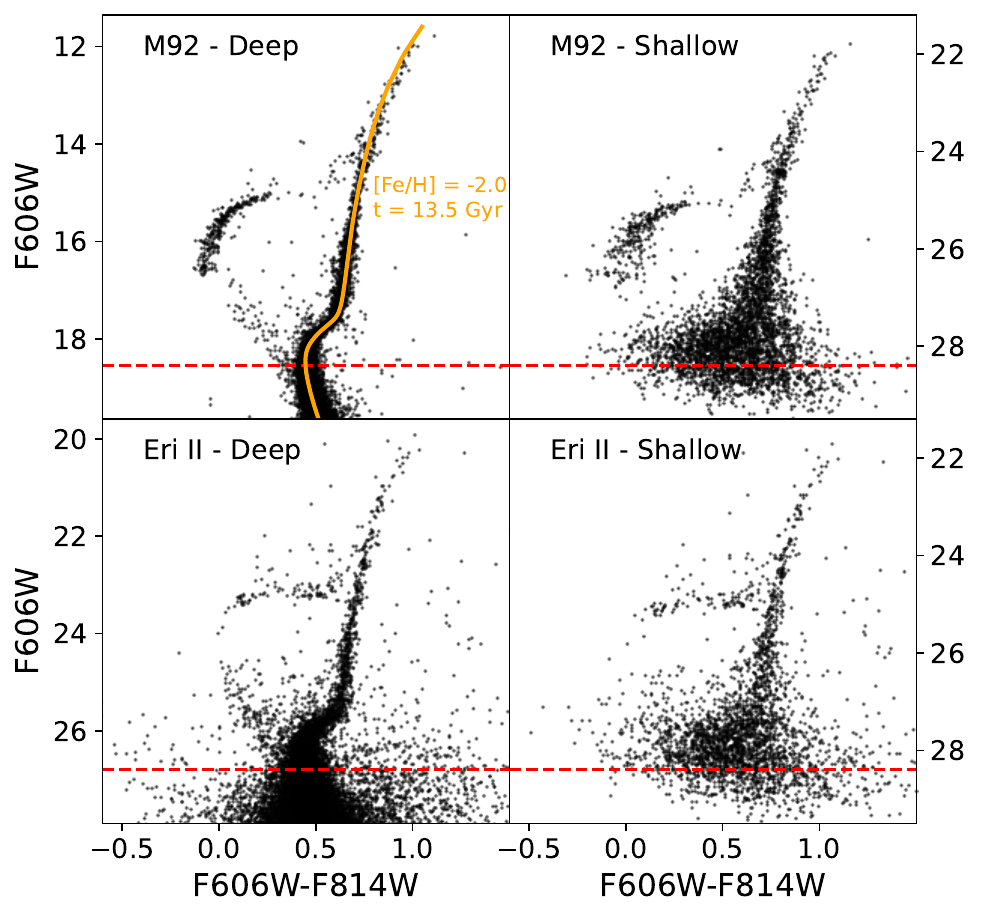}{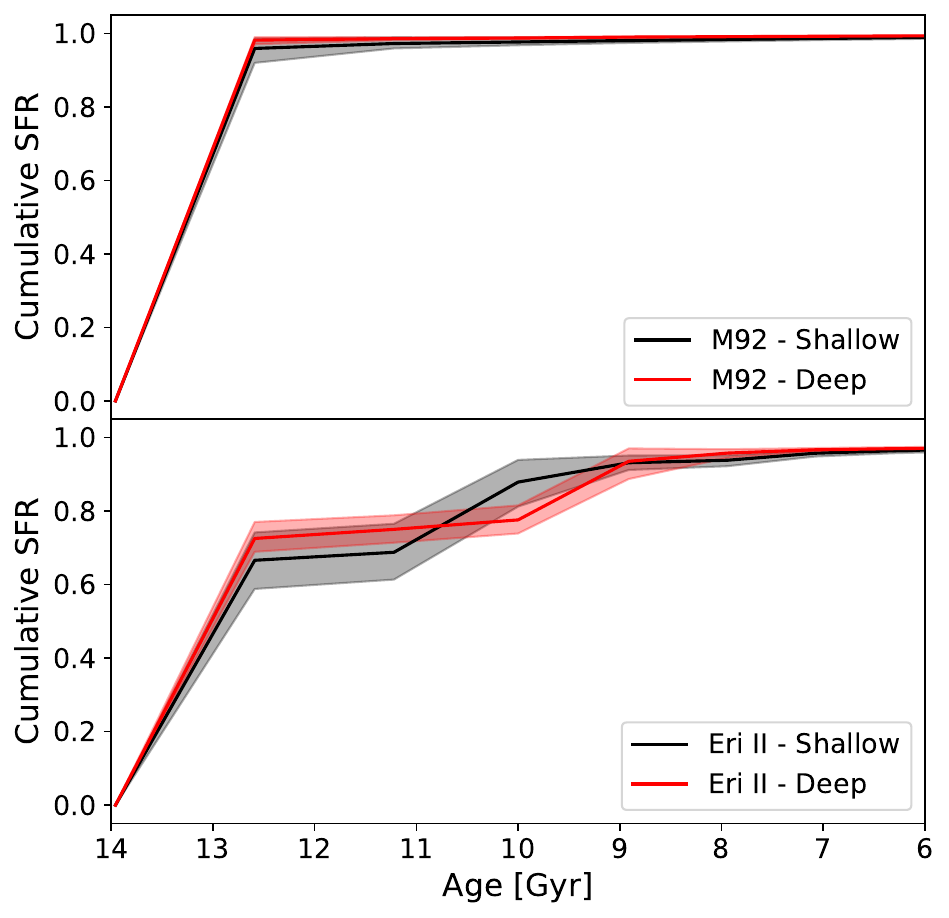}

\caption{Left panel: original CMDs of M92 and Eridanus II (left subpanels), obtained from deep \hst\ photometry, and noisier mocks (right subpanels) obtained by simulating the distance, reddening, completeness and photometric uncertainties of our target galaxies. The red dashed line shows the approximate location of the oMSTO in the different CMDs. A BaSTI isochrone, of the best-fit age and metallicity values, is superimposed on M92 (orange line) to illustrate the fidelity of our stellar models and the accuracy of the fit. Right panel: comparison the between the SFHs obtained from the deep photometry (red lines) and those obtained from the noisy photometry (black lines). The shaded regions represent the random uncertainties of our solutions.}
\label{fig:Depth}
\end{figure*}

\section{The effect of photometric depth}
\label{App:depth}
A clear difference between our observations of M31 satellites and the data used to derive the SFH of MW UFDs is the photometric depth. Due to the distance of M31, our CMDs are shallower and of lower S/N at the MSTO than the exquisite CMDs that have been obtained within the halo of the MW. To test how this difference may affect SFH recovery, we have taken two well-studied nearby stellar systems and created mock observations as if they were at the distance of M31. 

The results of this test are shown in Fig.~\ref{fig:Depth}. We begin with deep \hst\ observations of the globular cluster M92 (GO-9453, PI: Brown; GO-10775, PI: Sarajedini) and of the UFD Eridanus II (GO-14234, PI: Simon). These data have S/N values at the oMSTO of $\sim$600 (M92) and $\sim$40 (Eridanus II), which are both much larger than those of our M31 photometry ($S/N\sim10$). For this test, we first derived the CMD-based SFHs using the original photometry of M92 and Eridanus II. This is done using the same methodology outlined in the paper. We assume $(m-M)_0 = 14.59$ and E(B-V)=0.02 for M92 \citep{Harris10}, and $(m-M)_0 = 22.84$ and E(B-V)=0.01 for Eridanus II \citep{Martinez-Vazquez21}. The resulting SFHs are the red lines in the right panel of Fig.~\ref{fig:Depth} and will be assumed as our ground truth. We also use the well-known properties of M92 to further validate our overall methodology. The SFH we recover for M92 essentially corresponds to a simple stellar population with age and metallicity in excellent agreement with literature values \citep[e.g., ][the difference between our best-fit {$\rm [Fe/H]$} of -2.0 and the measured abundance of -2.35 is completely due to our use of scaled-Solar models to fit an $\alpha$-enhanced population]{Dotter07,Harris10, Vandenberg13}. The corresponding best-fit isochrone is also shown, in Fig.~\ref{fig:Depth}, on the M92 CMD to illustrate the fidelity of the BaSTI stellar models in the F606W/F814W photometric bands.

We then shifted the photometry of these two targets to simulate the same distance and reddening of one of our galaxies (\A{XIII}), i.e., $(m-M)_0 = 24.45$ and $E(B-V)=0.14$.  We then perturbed the measurements using the photometric uncertainties and the completeness estimated from the artificial star tests of \A{XIII}. The resulting CMDs (left panel of Fig.~\ref{fig:Depth}), simulate how the M92 and Eridanus II CMDs would appear if they were satellites of M31 observed by our Treasury program. We repeat the CMD modeling for these noisy CMDs and find that, besides the larger uncertainties, the resulting SFHs (black lines in the right panel of Fig.~\ref{fig:Depth}) are virtually indistinguishable from the ones obtained from the deep CMDs. The main difference is the increase in uncertainty, most notable for Eridanus II, owing to the lower S/N at the MSTO. This test therefore rules out any appreciable systematic in the SFHs caused by the limited S/N of our data at the MSTO.

\section{The effect of CMD-fitting code}
\label{App:Code}
\begin{figure*}
\plotone{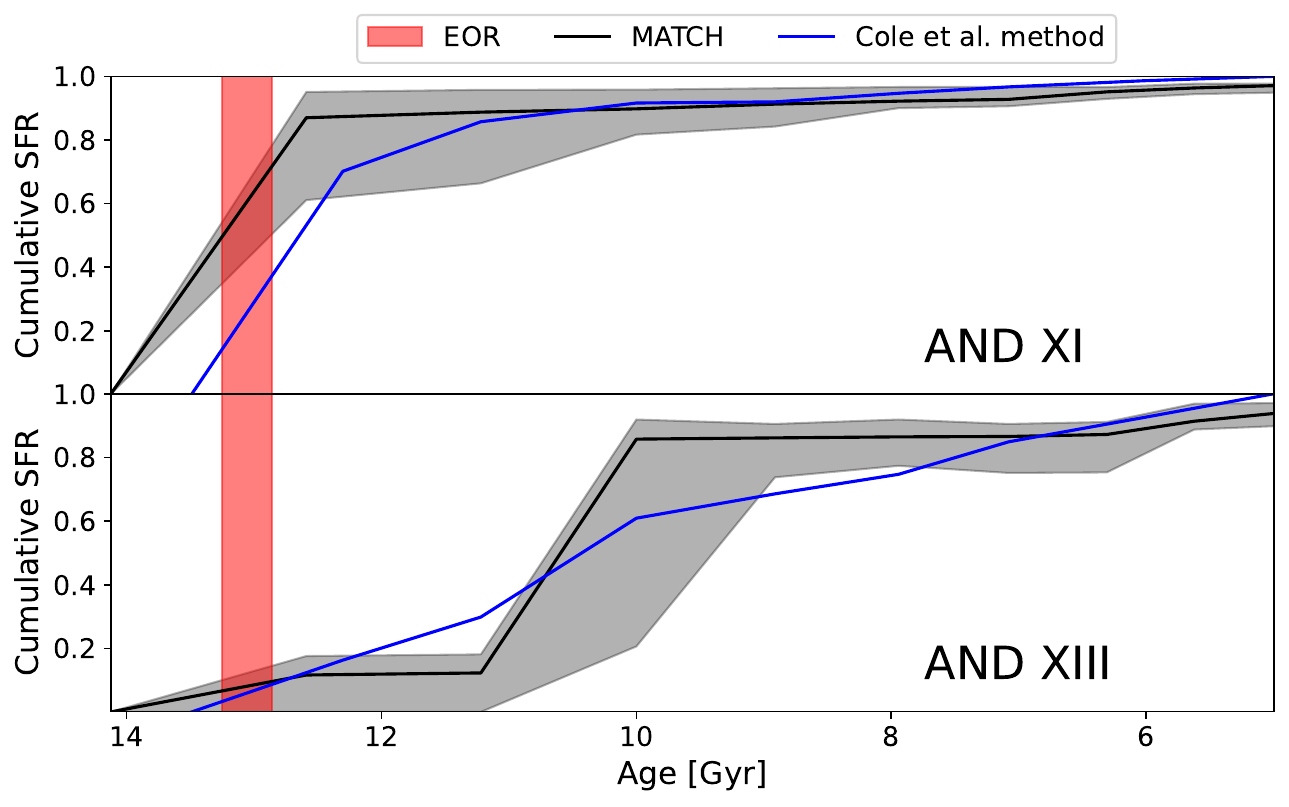}

\caption{Comparison between our fiducial SFH solution (black) and that obtained using the Cole methodology (blue), for \A{XI} (top panel) and \A{XIII} (bottom panel). The shaded grey region outlines the systematic uncertainties estimated through the method of \citet{Dolphin12}. The red shaded region marks the epoch of reionization.}
\label{fig:Code}
\end{figure*}

We checked the potential impact of the choice of CMD-fitting software on our derived SFHs by comparing our solutions to those obtained with an independent methodology. We use the method described in \citet{Cole07, Cole14} to re-derive the SFHs of \A{XI} and \A{XIII}, which have the oldest and youngest stellar populations in our sample, respectively. We use identical photometric samples, stellar libraries, distance, and reddening as used in the main analysis, to isolate the effect of the fitting methodology on the results. We summarize the setup of the ``Cole" method below.

The BaSTI isochrones at each metallicity from Z = 0.0001-0.001 are binned into uniformly logarithmically-spaced age groups from 9.70 $\leq \log(t) \leq$ 10.13 before being weighted by the initial mass function. Stars in each age bin are allowed to have any metallicity in the library, but the oldest bins ($\log(t) \geq 10.0$) are prevented from having Z $>$0.0006. Each age bin is initially taken to be 0.10~dex wide, but may grow or shrink during the fitting depending on the noise level in the fit. We take the initial-mass-function from \citet{Chabrier03}, which is log-normal within
0.08 $\leq M/M_{\sun} \leq$ 0.8. Only 35\% of the stars are assumed to be true singles and the rest are binaries; distinguishing between wide and close binaries, three-quarters of the binaries have the secondary mass drawn from the Chabrier initial-mass-function with the remainder drawn from a flat (uniform) initial-mass-function. The weighted isochrones are shifted by the adopted distance and reddening, convolved with the color and magnitude shifts from the artificial star tests, and weighted by the incompleteness fraction. The CMD is divided into bins that are 0.05 by 0.10 mag in size, and the SFH and AMR are derived by maximizing the Poisson likelihood for star counts in each bin that contains at least one star (excluding the same region of the HB and red clump that was excluded in the \texttt{MATCH} fits).

Figure~\ref{fig:Code} shows the results of the CMD fit, compared to the fiducial solutions, for \A{XI} and \A{XIII}. The Cole solution for \A{XI} confirms that the majority of stars in this galaxy are compatible with the oldest ages allowed by the model grid. The differences between the Cole and MATCH solutions is entirely attributable to the different extension of the grid to old ages. The star formation in \A{XIII} is distributed on a slightly longer episode in the Cole solution, compared to the narrower burst obtained with \texttt{MATCH}. Overall, the difference between the two \A{XIII} solutions is mostly contained within the systematic uncertainty envelope (shaded grey region) and confirms the broad consistency with \texttt{MATCH} in previous studies of isolated dwarf galaxies \citep[e.g.,][]{Monelli10a,Monelli10b,Skillman14}. In spite of these subtle differences, the comparison between \A{XI} and \A{XIII} is unchanged. The Cole SFH of \A{XI} reveals a primarily ancient stellar population, while \A{XIII} shows prominent star formation well beyond reionization, with only $\sim 15\%$ of the total star formation happening at $t<12.3$~Gyr ($z<4$). This test supports the robustness of our results against the choice of CMD-fitting methodology.



\end{document}